\documentclass[11pt,a4paper]{article}

\usepackage[font=it,labelfont=it]{caption}
\usepackage{makeidx}  
\usepackage{graphicx} 
\usepackage{a4wide}
\makeindex 

\begin{document} 

\title{Metal-loaded organic scintillators for neutrino physics}
\author{Christian Buck \\ Max-Planck-Institut f\"ur Kernphysik, 69117 Heidelberg, Germany \\ Minfang Yeh \\ Brookhaven National Laboratory, Upton, NY 11973, US}
\maketitle

\abstract{Organic liquid scintillators are used in many neutrino physics experiments of the past and present. In particular for low energy neutrinos when realtime and energy information are required, liquid scintillators have several advantages compared to other technologies. In many cases the organic liquid needs to be loaded with metal to enhance the neutrino signal over background events. Several metal loaded scintillators of the past suffered from chemical and optical instabilities, limiting the performance of these neutrino detectors. Different ways of metal loading are described in the article with a focus on recent techniques providing metal loaded scintillators that can be used under stable conditions for many years even in ton scale experiments. Applications of metal loaded scintillators in neutrino experiments are reviewed and the performance as well as the prospects of different scintillator types are compared.}

\newpage

\tableofcontents\clearpage

\section{Introduction}
Neutrinos constitute a fundamental ingredient in the Standard Model of particle physics and their properties are highly relevant in cosmology. In the last twenty years neutrino flavor oscillations were discovered in several experiments, implying that neutrinos are massive particles. With the improved knowledge about neutrino parameters we can now use current and upcoming neutrino detectors to learn more about their sources such as the Sun, the Earth, supernovae or nuclear reactors. Moreover, there is still room for additional discoveries and physics beyond the Standard Model in this field.  

Metal loaded organic liquid scintillators (LS) have been used in neutrino detectors from the first neutrino experiment of Reines and Cowan~\cite{R53}. They searched for neutrino interactions in a cadmium loaded LS. Since then, many other neutrino experiments have counted on the advantages of the LS technology. Some of them prefered metal loaded versions, others used liquids without metal loadings. The advantages of such a detector type are amongst others high purity, low energy threshold, detector homogeneity, flexible handling, scalability to large volumes and rather low cost.   

The main challenge in metal loading is to reach the solubility requirements which can vary from 0.1-10~wt.\% (depending on the application) without degrading the optical properties of the liquid. A polar material (metal salt) has to be dissolved in a non-polar liquid (scintillator solvent). Moreover, the liquid has to be optically and chemically stable over several years, the typical timescale for neutrino experiments. Such time scales are needed to collect sufficient statistics given the low neutrino interaction cross sections and associated low count rates. Huge progress has been made, in particular during the last 20 years, to meet the goals stated above. This article summarizes the lessons learned from pioneering experiments and the efforts made to overcome the difficulties faced in the past. In some applications scintillator loading of non-metal elements such as noble gases is required. The general challenge in scintillator loading is to maintain the radiopurity levels required in neutrino detectors and not to degrade the optical properties of the liquid. 

Chapter 2 starts with a description of the mechanism of light production in LS in general and discusses quenching effects. Basic scintillator components are introduced. The understanding of energy transfer processes between scintillator molecules, radiative and non-radiative, is essential for good scintillator design. Radiopurity and optical transparency of these components are a prerequisite for detector liquids in the search for solar, long and short baseline reactor, $\beta\beta$-decay or geo neutrino experiments. Sophisticated purification techniques were developed and applied to reach outstanding purity levels never observed in other fields. The most important properties and requirements of scintillator mixtures are described and compared for different compositions. 

In chapter 3 techniques of metal loadings into the organic liquid are described and the impact on the LS properties is characterized. We focus in more detail on three loading methods. One is the use of carboxylic acids. This approach was already used in the early stages by Reines and Cowan in the 1950s and then continuously further developed and optimized for several metallic elements. An alternative method is to achieve the solubility by dissolving the material in the form of metal-$\beta$-diketonates. More recently, a new development has started, which involves quantum dots with the prospect of adding the metal in form of a tunable fluorescent nanocrystal. Another promising concept is to make use of water based detectors with metal loadings, not only for Cherenkov, but also for scintillation light.  

In chapter 4, the performance of different metal loaded LS in various neutrino experiments is summarized. In reactor neutrino experiments loadings of gadolinium  (Gd), cadmium (Cd) or lithium (Li) were used to provide better neutrino detection and reduce background. Several experiments reported instabilities of their LS, sometimes limiting the precision of the physics results. Recent reactor neutrino experiments demonstrated that Gd-loaded LS can be used under stable conditions for neutrino detection even after 5 years of LS production and beyond. For solar neutrino experiments there was an interest in heavily loaded indium (In) and ytterbium (Yb) LS. Extensive Research and Development work was done and some conclusions of this effort are reported here. In projects looking for neutrinoless $\beta\beta$-decay LS loaded with e.g.~neodynium (Nd) and tellerium (Te) were studied. The chapter closes with an outlook and perspectives for upcoming experiments using LS technologies. 

\newpage
\section{Organic liquid scintillators}
\subsection{Mechanism of scintillation light production}
Light production in liquid scintillators (LS) involves many different processes and complicated microphysics on molecular scales. Part of the energy from ionizing particles passing through fluorescent organic materials is transformed into excitation of the aromatic solvent molecules. Excitation happens mainly for the delocalized electrons in the $\pi$-bonds of phenyl groups. After excitation radiationless conversion processes lead to population of the first excited electronic state. The time scale of these processes is in the order of $10^{-12}-10^{-11}$s~\cite{Bir}.

\begin{figure}
\begin{center}
\includegraphics[scale=0.15]{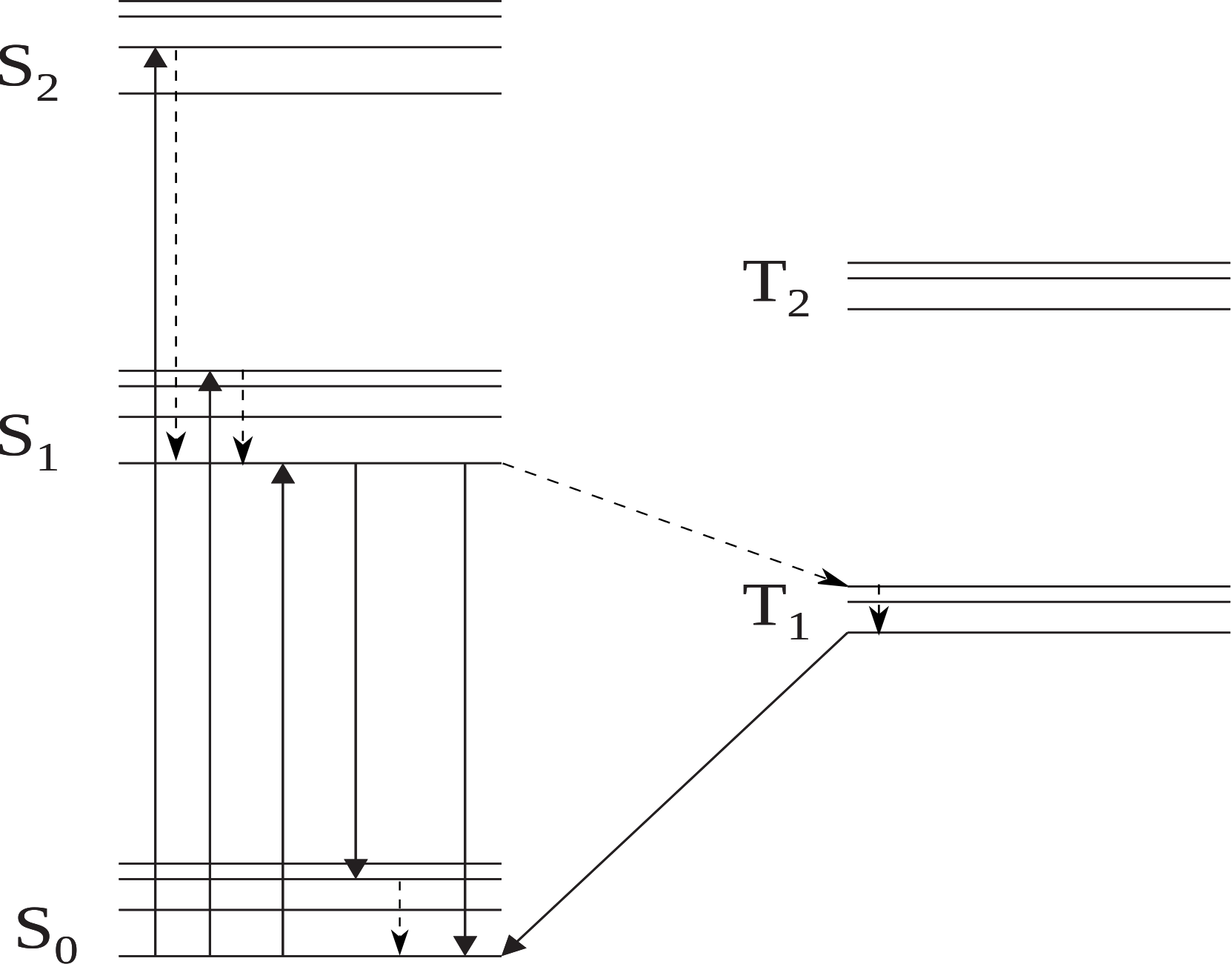}
\end{center}
\caption{The energy levels of singlet (S) and triplet (T) states with the
corresponding vibrational levels are shown for a fluorescent molecule. Solid arrows mark transitions leading to fluorescence or phosphorescence radiation whereas dotted arrows represent radiationless transitions.}
\label{Fig1}
\end{figure}

In an isolated fluorescent molecule the transition from the first excited singlet state to the ground state occurs via the emission of a photon. Such a fluorescence process happens within few to tens of nanoseconds. On longer time scales (milliseconds or longer) transitions via triplet states are also possible (phosphorescence) as depicted in Figure~\ref{Fig1}. In an organic liquid with just one aromatic solvent molecule (see section~\ref{secsol}) most of the floursecent radiation would be self-absorbed after short distance due to a significant overlap of absorption and emission spectra. The quantum yield of typical solvents is less than 50\%. Therefore, the pure solvent would be opaque for scintillation light at the dimensions of neutrino detectors. 

This is different for scintillation detectors using liquid noble gases. In liquid argon or liquid xenon for example, scintillation light is produced via the formation of excited dimer molecules. The excitation levels of those excimers do not correspond to the levels of the single atoms in the liquid noble gas. Losses due to self-absorption of scintillation light in the medium are thus reduced in such systems.

To get the scintillation photons out in an organic liquid system, an additional component has to be added with a good overlap of its absorption peak with the emission spectrum of the solvent molecules. This additional fluorescence molecule is typically called the ``fluor'' or  ``wavelength shifter'' (WLS) and has the purpose of shifting the emission spectrum to longer wavelengths into more transparent regions of the LS. Energy transfer from the solvent to the fluor is mainly radiationless, e.g.~via the F\"orster resonance energy transfer (FRET) mechansim~\cite{Fo48}. Scintillation light is emitted isotropically from the point of production.

Besides scintillation light production there is also the possibility of Cerenkov light creation. Cerenkov light is emitted when the velocity of a charged particle exceeds the speed of light in the medium with the refractive index $n$. Typical values for $n$ in a LS are around 1.4-1.6 in the optical blue region. The angle of the Cerenkov light cone depends on the particle kinetic energy and $n$, and gets narrower with higher velocities of the particle. Part of the prompt Cerenkov light in the UV region is absorbed by the scintillator material and reemitted at the emission spectrum of the fluors. In addition, there is a blue component of the Cerenkov light, which is in the transparent region of the LS. For electrons in the MeV range the Cerenkov contribution to the light production is typically an order of magnitude lower than the one of scintillation light.

\subsection{Quenching}
Ideally, excitation and ionization of the scintillator molecules along the path of the incident particle are separated by several molecular distances and interactions between excited molecules can be neglected. In such a scenario, the energy emitted as scintillation light ($L$) is proportional to the particle energy ($E$). In reality those assumptions are not always fulfilled. Losses of scintillation light are called ``quenching''. These losses can be particle and energy dependent leading to partly non-linear behaviour in the scintillation response.

The most prominent form of this effect is known as ``ionization quenching'' defined as the loss of scintillation light attributed to quenching of the primary excitation caused by a high density of ionized and excited molecules~\cite{Bir}. The quenching is strongest in case the specific energy loss dE/dx of the ionizing particle in the LS is large. This behaviour is observed around the end of a particle track or for low energies. The quenching is stronger for heavy particles such as in alpha radiation or in proton recoils as compared to electrons, which have a lower dE/dx. The detector response to alpha particles can be more than a factor 10 lower compared to electrons of the same energy. For a high ionization density more triplet states are excited. This can change the decay times of the scintillator, opening the possibility of particle identification in the LS via pulse shape discrimination (PSD) methods. 

Most experiments use the basic formula of Birks'~\cite{Bir} to describe ionization quenching:
\begin{equation}
\frac{dL}{dx}=\frac{L_0 \, \left(\frac{dE}{dx}\right)}{1+kB\left(\frac{dE}{dx}\right)}\,.
\end{equation}
Here, $dL$ corresponds to the light produced by a particle within a distance $dx$, $L_0$ is the light yield constant, and $kB$ is the quenching parameter. For electrons typical values of kB are in a range between 0.01-0.03~cm/MeV~\cite{Wag}. Although this equation is widely used, it should be emphasized that it is an approximation originally developed for inorganic crystal scintillators. If higher precision is desired a more  generalized form with two quenching parameters $A$ and $B$ can be used~\cite{Chou}:
\begin{equation}
\frac{dL}{dx}=\frac{L_0 \, \left(\frac{dE}{dx}\right)}{1+A\left(\frac{dE}{dx}\right)+B\left(\frac{dE}{dx}\right)^2}\,.
\end{equation}
There are also other more detailed models by Wright~\cite{Wri} or Voltz et al.~\cite{Vol} which try to take into account the complex processes of the quenching mechanisms on a molecular level. There can be relevant effects related to the production of secondary electrons by the primary particle or the involvement of triplet states. In the case of electrons in a LS, the ionization quenching at low energies (few MeV) can be reasonably described by all of the above models to first approximation~\cite{Wag}. However, there can be significant deviations in other situations, e.g.~for higher charges of the primary particle.  

Another type of quenching is the so-called ``impurity quenching'' e.g.~due to oxygen dissolved in the scintillator. In this case part of the energy of the excited solvent molecules is transferred to the non-fluorescent impurities and lost for scintillation.  

\subsection{Scintillator components} 
\subsubsection{Solvents}
\label{secsol}
The solvent molecules are the basis of each organic LS. Most solvent candidates are benzene derivatives, in particular alkyl benzenes like toluene, xylene or cumene, with an absorption band around 260~nm. Some basic properties of commonly used molecules are listed in Table~\ref{solvents}. The solvent of choice for high light yield scintillators in past neutrino experiments was often pseudocumene (PC, 1,2,4-trimethylbenzene). However, over time safety considerations became more and more important and the focus shifted to high flash point solvents as linear alkyl benzene (LAB), phenyl xylylethane (PXE) or di-isopropyl naphthalene (DIN). 

Material compatibility arguments sometimes exclude the use of pure PC or other chemically aggressive solvents. For example, when acrylic material is used as a detector vessel, the aromatic solvent is sometimes diluted in mineral oil (e.g.~medicinal white oil) or n-alkanes, which can also help to raise the flash point of the scintillator cocktail. There can be the difficulty of electrostatic charging when the solvent is passing through pipes and filters, in particular when these materials are not conductive. Therefore, in some cases, to improve fire safety, an anti-static agent might be added to the solvent as e.g.~Stadis-425 in the LS of the NOvA experiment~\cite{NOvA}. 
 
\begin{table}[h]
\caption[Solvents]{Density, flash point and the wavelengths of the optical absorption/emission peaks (dissolved in cyclohexane) for several solvent candidates are shown. \label{solvents}}
\begin{center}
\begin{tabular}{lccrcc}
Molecule & chemical formula & density [kg/l] & flash point & abs.~max. & em.~max. \\
\hline
PC & C$_9$H$_{12}$ & 0.88 & 48$^\circ$C & 267~nm & 290~nm \\
toluene & C$_7$H$_8$ & 0.87 & 4$^\circ$C & 262~nm & 290~nm\\
anisole & C$_7$H$_8$O & 0.99 & 43$^\circ$C & 271~nm & 293~nm\\
LAB & -- & 0.87 & $\sim140^\circ$C & 260~nm & 284~nm \\
DIN & C$_{16}$H$_{20}$ & 0.96 & $>140^\circ$C & 279~nm & 338~nm \\ 
o-PXE & C$_{16}$H$_{18}$ & 0.99 & 167$^\circ$C & 269~nm & 290~nm \\
n-dodecane & C$_{12}$H$_{26}$ & 0.75 & 71$^\circ$C & -- & -- \\
mineral oil & -- & 0.82 -- 0.88 & $>130^\circ$C & -- & -- \\
\hline
\end{tabular}
\end{center}
\end{table}

A solvent candidate becoming very popular in the last decade is LAB, mainly due to its superior safety features, good material compatibility, high transparency and low cost. It was first proposed in the context of the SNO+ experiment~\cite{Chen}. On the other hand LAB based scintillators have reduced pulse shape discrimination capabilities and as a mixture of many isomers it does not have a well defined hydrogen or carbon fraction. 

\subsubsection{Wavelength shifters}
For some applications it is sufficient to have just one wavelength shifter (WLS, fluor) in the scintillator mixture. This dopand is typically added at concentrations of few g/l. An additional secondary WLS can be advantageous to improve the attenuation length of the scintillator. In particular if the detector volume is very large or the absorption band of the solvent molecule has a long tail towards longer wavelengths, an additional WLS is added at concentrations in the order of 20~mg/l or less. 

\begin{table}[h]
\caption[Fluors]{The wavelengths of the optical absorption and emission peak maxima for primary and secondary fluor molecules diluted in cyclohexane are listed. \label{fluors}}
\begin{center}
\begin{tabular}{llcccc}
Molecule & chemical formula & abs.~max. & em.~max.\\
\hline
PPO	& C$_{15}$H$_{11}$NO & 303~nm & 358~nm \\ 
PBD & C$_{20}$H$_{14}$N$_2$O & 302~nm & 358~nm \\
butyl-PBD & C$_{24}$H$_{22}$N$_2$O & 302~nm & 361~nm \\ 
BPO & C$_{21}$H$_{15}$NO & 320~nm & 384~nm \\
p-TP & C$_{18}$H$_{14}$ & 276~nm & 338~nm \\
TBP & C$_{28}$H$_{22}$ & 347~nm & 455~nm\\
bis-MSB & C$_{24}$H$_{22}$ & 345~nm & 418~nm \\
POPOP & C$_{24}$H$_{16}$N$_2$O$_2$ & 360~nm & 411~nm \\
PMP & C$_{18}$H$_{20}$N$_2$ & 295~nm & 425~nm \\
\hline
\end{tabular}
\end{center}
\end{table}

Table~\ref{fluors} shows the wavelengths of maximal absorption and emission for primary and secondary fluors. To minimize losses and ensure efficient energy transfer between the molecules, a large overlap between the emission spectrum of the donor molecule and the absorption band of the acceptor is mandatory. In general the absorption and emission spectra shift to longer wavelengths as more aromatic rings are added to the molecule. 

A commonly used fluor combination is 2,5-Diphenyloxazole (PPO) and 4-bis-(2-Methyl-styryl)benzene (bis-MSB), although a better overlap of spectra could be reached using p-terphenyl (p-TP) with bis-MSB. However, p-TP has limited solubility in many solvents. Other alternatives for the primary fluor with very similar optical properties as PPO are 2-(4-biphenyl)-5-phenyl-1,3,4-oxadiazole (PBD) and 2-(4-biphenyl)-5-(4-tert-butyl-phenyl)-1,3,4-oxadiazole (butyl-PBD). The fluor 2-(4-biphenyl)-5-phenyloxazole (BPO) is known to provide high light yields, but on the other hand it is less favored due to its limited transparency in the region of scintillator emission and its toxicity. 

As secondary WLS 1,4-bis(5-phenyloxazol-2-yl)-benzene (POPOP) can also be used, which has a slightly longer decay time than bis-MSB. In several liquid argon scintillation detectors tetraphenyl-butadiene (TPB) coated films are preferred for efficient light conversion from the argon emission (128~nm) to the blue-visible region (400--500~nm), where photomultiplier tubes (PMTs) are most sensitive. An exceptional large Stokes shift, the difference between positions of the absorption and emission peak, is observed in 1-phenyl-3-mesityl-2-pyrazoline (PMP). The PMP Stokes shift is about twice as large than in other fluor candidates. 

A common feature of all WLS is the high fluorescence quantum yield (QY), the probability to reemit a photon after absorption. When diluted in an inert medium as cyclohexane the QY of basically all molecules of Table~\ref{fluors} is above 80\%~\cite{Ber, QY}. However, there can be strong concentration and solvent effects changing the QY numbers significantly. Moreover, for many fluors there are contradictory results in the literature values of this quantity~\cite{Hein}.

\subsection{Energy transfer}
Ionizing radiation in a LS excites the solvent molecules. This excitation energy should then be converted into scintillation light, which can be detected by the photosensors. The energy transfer from the solvent to the fluor in a LS is mostly non-radiative. The transfer can be via collisions between a donor molecule D and an acceptor molecule A, but also other processes are relevant. The radiationless transfer can also occur over distances much longer than the dimension of the corresponding molecules. Several general models exist to describe the energy transfer between organic molecules. The models described below are sometimes applied for the specific case of LS. 

In a simple model by Perrin~\cite{Per} the energy is transferred from D to A, if the distance $R$ between the molecules is smaller than a critical distance $R_0$. The efficiency $\eta$ depends on the acceptor concentration $c$ and the critical concentration $c_0$:

\begin{equation}
\eta=1-e^{-\frac{c}{c_0}}
\end{equation}

In a kinetic approach of Stern--Volmer~\cite{Ste} the transition probability is independent of the distance between A and D and the efficiency follows

\begin{equation}
\eta=\frac{1}{1+\frac{c_0}{c}}
\end{equation}

A quantum mechanical treatment of F\"orster~\cite{Fo48} assumes D and A are isolated or at least diluted in an inert medium, which is transparent in the wavelength region of interest. Moreover, the distances between the molecules are assumed to be large compared to their dimensions. In the F\"orster model the energy transfer is based on interactions of the electrical dipole fields of D and A. Here, the transition probability strongly depends on the distance $R$ between the molecules and is proportional to a factor 1/$R^6$. The transfer efficiency can be written as~\cite{Fo59}

\begin{equation}
\eta=\sqrt{\pi}\cdot x\cdot \exp(x^2)[1-erf(x)]
\end{equation}

\noindent with $x=\frac{1}{2}\sqrt{\pi}\frac{c}{c_0}$ and the error function $erf(x):=\frac{2}{\sqrt{\pi}}\int_0^x{e^{-u^2}du}$.

The critical transfer distance $R_0$ is defined as the distance for which the probability of spontaneous decay with photon emission and energy transfer from D to A are the same. It depends on the quantum yield of D and the overlap of the donor emission and acceptor molar extinction spectrum. For an efficient transfer a large overlap between the spectra is desired. Typical values of $R_0$ are in the few nanometers range. At the critical concentration $c_0$, which is proportional to $\frac{1}{R_0^3}$, $\eta$ is about 0.7.

The model by Dexter~\cite{Dex} is an extension of the one by F\"orster. In addition to the resonant energy transfer between allowed transitions, it includes interactions involving quadrupole fields or exchange interaction. Energy transfer via the exchange interaction dominates at short distances of less than 1~nm. The transition probability is proportional to $\exp(-2R/L)$, where $L$ is a positive constant. As for the other energy transfer processes, the transfer rate of the exchange interaction is most efficient for a large overlap between absorption and emission spectra of the acceptor and donor molecules. 

In organic LS there are effects of major relevance, which are not included in the above models. These effects are related to the high concentrations of solvent and primary fluor molecules. Models have been developed trying to include them for more realistic descriptions~\cite{Kal}. Since the required distance for efficient energy transfer from the excited donor (solvent molecule) to the acceptor (fluor molecule) can be larger than the solvent to fluor distance, there is significant donor to donor energy transfer first. Instead of being transferred in one step from solvent to fluor, the excitation energy can be passed between neighboring donors before reaching the vicinity of an acceptor (migration transfer or ``hopping''). Moreover as secondary effect, in a liquid, in contrast to a solid material, the excited molecules can move e.g.~by diffusion and in this way change the distance to the acceptor. Therefore light production in a has some dependence on viscosity and temperature of the fluid. The probability for an excited solvent molecule to decay via the emission of a real photon, which escapes from the solution can be neglected in most systems. 

Radiative energy transfer is more relevant for the light propagation in a LS after primary scintillation light production. The light emitted by the fluor undergoes absorption and re-emission processes. Therefore, it is very important to have a high fluorescence quantum yield for the fluor molecules, which determines the re-emission probability.

\subsection{Purification techniques}
Experiments like Borexino~\cite{Bor} and KamLAND~\cite{Kam} have demonstrated the capabilities of LS in experiments with very low count rates in the MeV region. Several different purification techniques are available to improve the performance of a LS detector. An essential step is nitrogen purging to remove spurious gases dissolved in the liquid such as oxygen, which has a quenching effect on the scintillator light yield. Efficient methods that are used to remove other optical or radio-impurities are distillation, water extraction or column purification, sometimes also combinations of them.

In a neutrino detector impurities below the sensitivity of chemical analysis techniques in the laboratory can still fail to satisfy the specifications for the radiopurity (e.g.~$<$~10$^{-16}$~g/g uranium or thorium concentrations) or optical transparency ($> 10$~m in the 400--500~nm region) constraints. Sometimes, the specific molecular form of the impurities is even not exactly known. The purification methods applied should try to be as efficient as possible in the removal of potential trace contaminations. 

\subsubsection{Nitrogen purging}
Volatile components dissolved in the liquid as oxygen or radioactive noble gases are removed by gas, typically nitrogen, stripping. The method is based on differences in the equilibrium composition between liquid and vapor. Approximations on the purification factor can be made by use of Henry's Law. The technique is very effective in particular by applying counter current flow of the liquid and the gas. The efficiency can be further increased by using columns with structured packing~\cite{Ben} or higher temperature, implying reduced solubility of gas contaminants. To avoid impurities such as radon or krypton being added by the gas purging, the purification and selection of clean nitrogen is essential~\cite{Sim}.

\subsubsection{Distillation}
Distillation is an effective process to improve the scintillator transparency and to  reduce radioactive impurities in the solvent liquid such as daughters of the radon decay chain~\cite{Kee}. All impurities which are less or more volatile than the solvent can be separated. Since the more volatile components are effectively removed by gas stripping, the focus is usually more on the less volatile ones. When distillation is applied to the full scintillator mixture one has to be careful not to separate the solvent from the fluor~\cite{Ben}, since both of them have very different boiling points. Scintillator solvents are often extracted from oil products and separated from other oil components by distillation at the supply company. Mono-molecular solvents such as PC can be made intrinsically purer as mixtures, since the distillation parameter ranges for temperature and pressure can be set narrower.  

In particular, multi-staged vacuum distillation~\cite{Ford} is applied as a purification step for LAB and PC. The vacuum distillation column should be composed of many theoretical plates that can effectively separate the organic solvents with different boiling points, e.g.~at $\sim10^\circ$C intervals. Typically, distillation is set to finish when more than 90\% of the solvent is collected. The impurities could be for example partially oxidized organic molecules having lower volatility compared to the pure solvent. Distillation is one of the best methods for improving the optical transmission of a LS solvent~\cite{Ford}. Optical purity is essential for high light collection and to achieve good energy resolution in a neutrino detector.  

\subsubsection{Water extraction}
In the water extraction method impurities in the organic LS are transferred into an immiscible aqueous phase based on their relative solubilities. An advantage of water extraction over distillation is the ability to directly process the full scintillator mixture allowing for on-line repurification of the detector liquid. This method is highly effective at removing polar or charged substances. It significantly lowers the amount of radioactive impurities such as $^{238}$U, $^{232}$Th or $^{40}$K. Most of the radioimpurities typically enter the scintillator mixture via the primary fluor. In particular the potassium (K) content in a primary fluor as PPO can be in the ppm level, far above the usual specifications of a low background detector. Water extraction in concentrated PPO solutions was found to be a promising way of removing K~\cite{CTF}. On the other hand, water extraction is less suited reducing optical impurities, which are mainly of organic type and would require phase separation from the scintillator solvent.

\subsubsection{Column purification}
Traces of chemical impurities can be removed from the scintillator by sending the liquid through a solid column of an adsorber material as aluminum oxide, silica gel or an other molecular sieve. A prerequisite for a good adsorber material is a large surface-to-volume ratio. Other important properties are adsorber parameters such as pore size and surface conditions. Many adsorber materials are hygroscopic and should be dried at elevated temperature before use to remove water molecules blocking adsorption sites. The ratio of impurities on the adsorber surface and in the liquid is determined by a specific equilibrium constant. The equilibrium constant depends on the free enthalpy of the system and the temperature. The free enthalpy is constant for a given solvent, impurity and adsorber type.

At the beginning of the purification process the impurity concentration forms a gradient along the column from low purity at the entrance of the column and high purity at the end. At some point the impurity is travelling through the column and reaches the exit, the so-called breakthrough. In a dynamic situation of a constant flow, the number of segments in a column depend on the time constant to reach the equilibrium situation and the velocity of the liquid through the column.

A weak acidic alumina column was found to significantly improve the attenuation length of scintillator solvents as e.g.~o-PXE~\cite{PXE}. The purification normally has no significant effect on the scintillation yield. Since the chemical composition of most of the different impurities that are removed by the column are not well defined, several layers of activated column packings differing in pH can be used~\cite{LSP}. In this multi-stage column approach acids, bases as well as neutralized particles can be effectively removed.

\subsection{Scintillator properties}
One of the most relevant scintillator properties in neutrino experiments are optical clarity, light yield, light emission spectrum, decay times and radiopurity. For the study and simulation of light propagation in the liquid it is important to know about scattering, re-emission probabilities, quantum yields and refractive index. Furthermore other characteristics have to be considered in choosing the adequate components for a given application as safety aspects, chemical stability, material compatibility H:C ratio and pulse shape discrimination capabilities. Finally, for detector construction parameters such as density, viscosity, cost and ease of production are relevant. In this section the general properties of commonly used LS are reported. A comparison is made for some combinations of solvent and fluor candidates introduced above. The effect of the fluor concentration on certain parameters is discussed.  

\subsubsection{Light emission spectrum}
The emission spectrum of a LS is determined by the fluors, since the de-excitation of the solvent molecules is mainly non-radiative. To measure the characteristic emission spectrum of a fluor, the molecule is diluted in an optically and chemically inert solvent as cyclohexane. However, in the scintillator solvent the spectrum can be shifted by up to 10~nm compared to isolated molecules. The interaction with neighboring solvent molecules might affect the levels of the electronic states especially for polar substances. In addition there can be concentration effects, in particular for the primary fluor which is typically added at few g/l, which might cause a distortion of the spectrum.

As the light propagates through the liquid in a large scale detector, it can be absorbed and re-emitted by the fluor. Therefore the emission spectrum changes and shifts to longer wavelengths, since the photons are only re-emitted at equal or lower energies. 

To simulate the detector response it is important to know the primary scintillation spectrum after non-radiative energy transfer including all solvent and concentration effects. To avoid shifts of the spectrum due to intrinsic absorption and reemssion processes, the primary light emission should be detected at the same side of the cell where excitation takes place (``front face geometry''). Then an optical model should be applied to take care of the light propagation in the detector from the point of production to the photosensor. The model should include the most relevant parameters as absorption properties of the components, re-emission spectra and probabilities, quantum yield of the fluors and scattering cross sections. 

\subsubsection{Transparency}
The intensity $I$ of a one-dimensional monochromatic light beam propagating through a LS decreases exponentially with the attenuation length $\Lambda$. If the attenuation length is measured using a standard UV photospectrometer, light can be lost either by absorption or by scattering. The absorbance $A$ after a distance $x$ is defined as $A(x)=\log_{10}(I(0)/I(x))$. The total absorbance can be calculated by summing up the absorbance contributions of the different LS components. Accordingly, the overall attenuation length $\Lambda_{LS}$ is obtained from the single contributions $\Lambda_{i}$ via

\begin{equation}
\label{sumatt}
\frac{1}{\Lambda_{LS}}=\frac{1}{\Lambda_1}+\frac{1}{\Lambda_2}+...
\end{equation}

Here the different terms can represent different processes as absorption or scattering. The formula can also be used in case the $\Lambda_{i}$ are determined separately for the different components in the mixture from individual measurements. 

The linearity between absorbance and concentration is known as the Beer-Lambert Law. The validity of this Law is in principle constrained to low concentrations. However, if we look at wavelength regions in the tail of the spectra the absorbance is mainly on impurities of the sample, which are low in concentration. Therefore, equation \ref{sumatt} can be used in the region of LS emission which is mostly in the high wavelength tail of the spectra. 

\begin{figure}
\begin{center}
\includegraphics[scale=0.50]{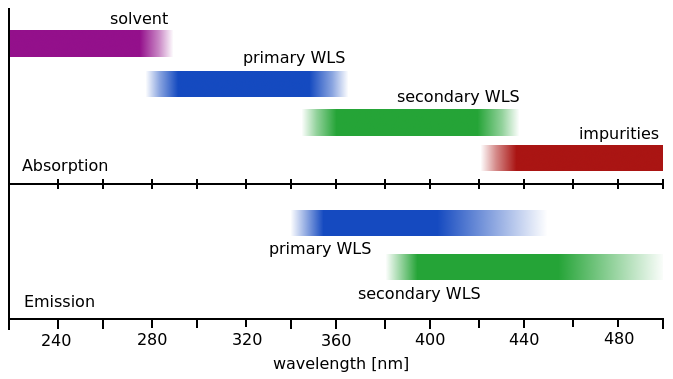}
\end{center}
\caption{The plot shows for each scintillator component the wavelength region for which its own absorbance is dominating in the liquid. In addition the typical wavelength ranges for scintillator emission is shown for the case of just a primary or including a secondary wavelength shifter.}
\label{Fig2}
\end{figure}

In the wavelength region of UV light below 280 nm the solvent molecules dominate the absorption in a LS as illustrated in Figure~\ref{Fig2}. At higher wavelengths, in the region of solvent emission around 300~nm, the main absorption is on the primary fluor, typically up to about 350~nm. The secondary fluor, if present, dominates absorption in the region of the scintillation light seen by the photomultipliers, which is normally above 350~nm. The molar extinction coefficient of secondary fluors like bis-MSB are dropping steeply above 400~nm and the absorbance starts to be negligible above 430~nm~\cite{LSP}. In this optical region the impurity levels in the chemicals and scattering processes are most relevant for the attenuation length. 

Highest attenuation lengths for solvents are typically found in pure n-alkanes as n-dodecane. As delivered from companies it can be already more than 10~m at wavelengths around 350~nm~\cite{LSP}. In typical cases the attenuation length is monotonically increasing from $350 - 580$~nm. Attenuation lengths of more than 10~m can also be found in LAB~\cite{Yeh2011}. For some LAB samples however, there can be wiggles in the absorption spectrum in the region between $350 - 400$~nm possibly caused by contaminations of bi-phenyls. In such a case a secondary fluor should be used. Pseudocumene has a slightly lower attenuation length above 400~nm than LAB or n-alkanes. For PXE or DIN the attenuation length is typically even lower, in a range of 1-5~m around 430~nm. However, the lower values of PC and PXE are partly due to a reduced scattering length compared to LAB or oil~\cite{WURM}. With alumina oxide column purification the attenuation could be improved to about 10~m for PXE~\cite{PXE} and to about 4~m for DIN~\cite{DIN}.       

\begin{figure}
\begin{center}
\includegraphics[scale=0.4]{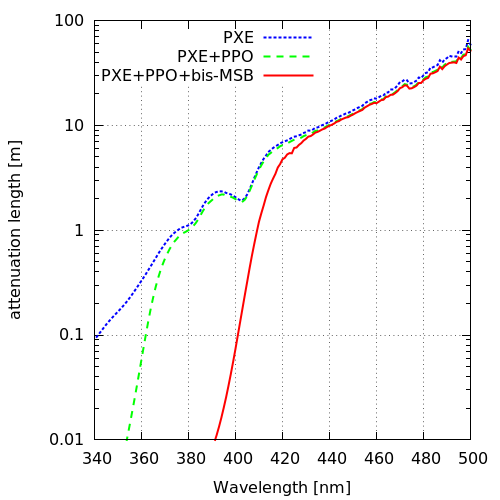}
\end{center}
\caption{Attenuation length of a pure PXE sample (dots), the same PXE with 3 g/l PPO (dashed) and after addition of 20 mg/l bis-MSB (solid).}
\label{Fig3}
\end{figure}

For high purity PPO the attenuation length in a transparent solvent at concentrations of few g/l is still well above 10~m~\cite{LSP}. In general, for solvents as well as fluors, there are significant batch to batch variations even for material from the same supplier and identical quality specifications. Therefore, careful material selection, testing, liquid handling as well as close contact to the companies are mandatory to reach the highest optical purity. As an example the attenuation length of purified PXE with and without fluors is shown in Figure~\ref{Fig3}.   

The nominal attenuation length of LS (after purification) can be longer than 10~m. The typical 10~cm pathlength UV spectrometry, which is adequate for long-term optical stability monitoring at 5\% variation, has limited precision for absolute measurements of the attenuation length at very high transmission. To further reduce the systematic errors, long-arm attenuation systems are essential. 

In a specific setup~\cite{Yeh2011}, a 2~m long stainless steel tube (inner) coated with Teflon is vertically aligned and contains several optical components in addition to a PMT. Such a system is optimized to measure long attenuation length LS. A cylindrical quartz tube (2.5~cm diameter) is connected to the steel tube where the liquid is contained for pathlength control. The vertical orientation allows the selection of various pathlengths in the liquid from a few centimeters to 2~m, in order to measure the variation of transmitted light as a function of propagation distance in the liquid. Pulsed LED beams, at selected wavelengths from 350-500~nm, are transmitted by an optical fiber to a beam-splitting cube that produces two beams with 50/50 intensity ratio. The different wavelengths give the capability to investigate the Rayleigh scattering, which has a sensitive dependence on wavelength. The split beams are collected by fiber collimators and transmitted again by 2~m and 40~m optical fibers, respectively, to give a time delay $\sim 200$~ns between the two pulses. One beam measures the absorption of the scintillator, while the other measures the variability of the pulsed LED output. Both light beams are collected by the same PMT and the signals are analyzed. Such a system gives a sensitivity of absorption length measurement at $>30$~m~\cite{Yeh2011}.

The transparency of liquid scintillators, whether they are metal loaded or metal free, is very sensitive to the liquid handling and storage conditions. The attenuation length of a pure PC solvent can reduce by a factor of more than three within few weeks when stored at room temperature in an open steel vessel~\cite{Ben}. Exposure of scintillator components to air can lead to oxidized products absorbing light in the region of scintillator emission. Materials such as stainless steel can act as a catalyst for the reactions and contact to the organic liquid might reduce the optical stability. To minimize such yellowing an antioxidant as Vitamin E or BHT might be added to the solution~\cite{NOvA}. It was found that Vitamin E is also fluorescent and can contribute to the scintillation~\cite{Mini}. 

\subsubsection{Scattering}
The attenuation length in a one-dimensional beam is limited by scattering. Whereas absorbed light with no re-emission is lost in a large scale LS detector, scattered light just changes propagation direction. Therefore, scattered or re-emitted light could still be detected at the photosensors of a three-dimensional detector. 

In a pure scintillator mixture Rayleigh scattering is dominating over Raman or Mie scattering. Without absorption, considering Rayleigh scattering alone the attenuation length at 430~nm for standard scintillator solvents is in the range 20-50~m~\cite{WURM}. This means for attenuation lengths of more than 20~m, as measured in a UV/Vis spectrometer, more light is scattered than absorbed without re-emission. In water based LS scattering can be significantly higher compared to pure organic liquids due to the organic/water interfaces. There is a strong wavelength dependence for Rayleigh scattering. The scattering length rises as $\lambda^4$. Although light scattering has a small impact on the energy resolution of a detector it has the negative aspect of deteriorating the capabilities of time or vertex reconstruction.

\subsubsection{Light yield}
In a two component scintillator system, for low fluor concentrations, the light yield (LY) strongly increases with increasing fluor loadings. Energy transfer from the solvent to the fluor competes with internal losses of the solvents. About half of the maximal LY is reached at the critical concentration~\cite{LYA}. This critical concentration depends on the solvent/fluor combinations and is lower if the non-radiative transfer is effective. The critical concentrations for standard scintillators is typically less than 1~g/l. 

\begin{figure}
\begin{center}
\includegraphics[scale=0.5]{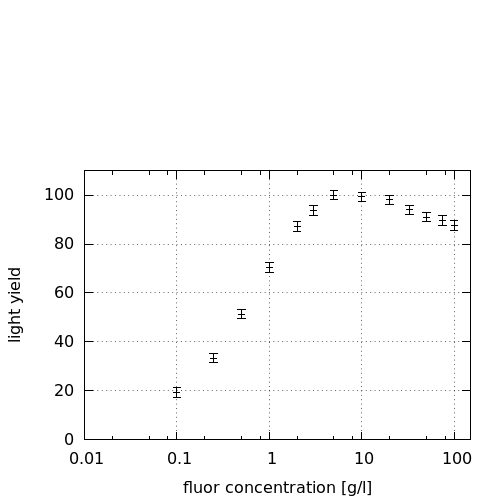}
\end{center}
\caption{Relative light yield of a scintillator with methoxybenzene as solvent and increasing BPO concentration. The values are normalized to the highest light yield measured at 5~g/l.}
\label{Fig4}
\end{figure}

Above the critical concentrations the LY keeps increasing with increasing fluor concentrations, but the curve flattens until it reaches saturation around $5-10$~g/l. At more than 10~g/l the light yield starts to lower again due to self-quenching. Self-quenching is an effect related to interactions between unexcited and excited fluor molecules in which the excitation energy can be lost by collision. The light yield dependence on the fluor concentration is illustrated in Figure~\ref{Fig4} for the case of BPO in methoxybenzene (anisole)~\cite{LYA}. 

Absolute measurements of the scintillation photons emitted for a given energy deposition in a scintillator liquid are difficult. Therefore most measurements are done relative to a reference scintillator with a known light yield. To minimize the systematic uncertainty samples and reference should be measured in the same setup. Typical reference samples of use are anthracene or the commercial scintillator BC-505 from Bicron company (St.~Gobain Crystals). In the data sheet provided by Bicron, the specified light yield of the BC-505 scintillator is 80\% anthracene. However, the literature values for the absolute light yield value for the anthracene itself (in photons per MeV) vary up to 25\%. 

Highest light yields of about 13000~photons/MeV are obtained with PC based scintillators. For Borexino, which is using pure PC as solvent, the yield is a little bit lower, 11500~photons/MeV~\cite{Eli}, since the PPO concentration is rather low (1.5~g/l). In KamLAND the PC is diluted in n-dodecane. The aromatic fraction is at 20\%, nevertheless the LY is still more than 80\% of the Borexino value at the same PPO concentration~\cite{Sue}. Even for pure mineral oil with a fluor there is some scintillation light. The amount depends on the exact chemical composition and in particular the aromatic fraction of the oil. The light yield with pure oil as solvent is usually less than 30\% of pure PC~\cite{Yeh2005}. But even for a pure n-alkane significant amount of scintillation light can be found~\cite{LYA}. In this case without aromatic ring structures and no $\pi$ electrons, the excited state is populated by processes like ionization or free radical formation with subsequent recombination.

In the Borexino CTF~\cite{CTF}, a PXE based scintillator was tested with p-TP as primary and bis-MSB as secondary fluor. The light yield was measured to be 93\% of the Borexino mixture~\cite{PXE}. Typical values for LAB based scintillators are slightly below the LY found in the KamLAND mixture. The Daya Bay experiment found similar values for light yield and attenuation length in their Gd-loaded and unloaded LAB based LS~\cite{DBLSP}. In Table~\ref{prop} those numbers along with some other properties of several LS are listed.

\begin{table}[h]
\caption[prop]{Selected properties of several scintillators as light yield (LY, given in \% of anthracene), attenuation length $\Lambda$, decay time and hydrogen (H) to carbon (C) ratio. \label{prop}}
\begin{center}
\begin{tabular}{lccccc}
Experiment & LS composition & LY [\%] & $\Lambda$~[m] & decay time & H/C \\
\hline
 & BC-505 (Bicron) & 80 & & 2.5~ns & 1.33 \\
Borexino~\cite{Bor} & PC+PPO (1.5~g/l) & 70 & 7~m (400~nm) & 3.5~ns & 1.33 \\
KamLAND~\cite{Sue} & PC+n-dodecane+PPO & 57 & 10~m (400~nm) & 6~ns & 1.97 \\
CTF~\cite{CTF} & PXE+p-TP & 65 & 10~m (430~nm) & 3.2~ns & 1.13 \\
Daya Bay~\cite{DBLSP} & LAB+PPO+bis-MSB & 50 & 14~m (430~nm) & 4.1~ns & 1.64\\
\hline
\end{tabular}
\end{center}
\end{table}

In case of metal loadings an additional energy transfer path opens from the solvent molecules to the metal compound which competes with the transfer to the fluor. In the scintillator design this transfer to the metal material has to be minimized compared to the transfer to the fluor, since the metal compound in general is non-fluorescent. Therefore, light transferred to the metal compound is lost for the scintillation process and reduces the LY (quenching). The LY in metal loaded LS can be improved by an increase of the primary fluor concentration~\cite{LYA}, which strengthens the energy transfer path from the solvent to the fluor, or by adding an additional WLS. However, at too high fluor concentrations the light loss due to the self-quenching effect will exceed the light gain due to more effective energy transfer from the solvent to the fluor. 

\subsubsection{Timing properties}
The decay time in a LS can have several components. The lifetimes of the different molecular excited states in the scintillator medium determine the pulse shape of the events. Therefore, experiments using pulse shape analysis to discriminate their background strongly depend on the values of their fluorescence decay-time constants. The time behaviour of the scintillator emission is often described by a multiexponential function with three or more decay constants.

The shortest decay constant (few ns) is attributed to the fluorescence decay time of the primary fluor, but it also includes the time for energy transfer from the solvent to the fluor, which might occur in several steps. Therefore, with increasing fluor concentration the decay time gets faster (more direct fluor excitation) until the decay constant approaches the intrinsic value of the isolated fluor molecule. This shortest decay constant is the dominant contribution to the decay profile. The longer components could arise from de-excitation of triplet states or other rare processes. Scintillator mixtures based on oil admixtures or LAB have longer decay constants than solvents as PC or PXE~\cite{time}. The effect of adding a secondary fluor as bisMSB to a mixture containing PPO has only a small effect on the decay times.

\subsubsection{Radiopurity}
At low energies up to about 200~keV, the background in an organic LS detector is limited by the $\beta$-decay of intrinsic $^{14}$C (E$_{\mathrm{max}}=156$~keV), a component which can not be chemically separated from the liquid. In the atmosphere the ratio of $^{14}$C/$^{12}$C is about $10^{-12}$~g/g. The production mechanism of $^{14}$C is cosmic ray interactions on nitrogen. In petrochemical organics, which are stored underground for millions of years, most of the original $^{14}$C (half-live of 5730~years) has decayed. In this way, the isotopic fraction can be reduced up to 6 orders of magnitude~\cite{Bortech}.

To reach and maintain high radiopurity in a LS, contact of the liquid with dust (U, Th and K contaminations) and air (Kr and Rn) have to be avoided. In Borexino~\cite{Bor} and KamLAND~\cite{Kam} radiopurity levels of $10^{-17}$~g/g and less for $^{238}$U and $^{232}$Th were reported. It was found to be hard to remove the alpha emitter $^{210}$Po, a daughter isotope of the $^{238}$U chain. It is generated via $\beta$-decay of $^{210}$Bi, which itself is created by the long-lived $^{210}$Pb (half live of 22.3~years). Alpha particles can be discriminated in LS with good particle identification via pulse shape analysis or they are even below energy threshold because of ionization quenching. Therefore, the contribution of $^{210}$Bi might be more critical as compared to $^{210}$Po in terms of background~\cite{Bor, Kam2}. Those isotopes at the end of the U decay series are often not in secular equilibrium with the rest of the chain. They can be supplied from the surface of the detector vessel and might not be intrinsic impurities of the LS itself~\cite{Kam2}. The equilibrium can also be broken at the level of radium or the mobile and omnipresent noble gas radon. For $^{40}$K levels of less than $10^{-16}$~g/g could be reached. To reach the purity requirements in the solutes, in particular for K, U and Th, is typically more challenging than for the solvents.  

Special emphasis in scintillator purification is on radioactive isotopes of noble gases. Those are removed by gas stripping using ultrapure nitrogen~\cite{Sim}. In the nitrogen plant emanation of Rn isopotes from materials in contact with the gas have to be minimized. In some applications, e.g.~in Borexino, purity levels of less than one $^{222}$Rn atom per m$^3$ of LS are required and were achieved~\cite{Bortech}. Reaching and maintaining such radiopurity levels is even more difficult when the LS is loaded with metals or other elements.  

\newpage
\section{Metal-loaded liquid scintillators}
\subsection{Overview}
In most general applications liquid scintillators are binary or ternery systems as described in the previous chapter and consist of organic components. However, in neutrino physics there is often the need for metal loading. The challenge in the development of metal loaded LS is the addition of the inorganic material, typically in the form of salt in the starting product, to the LS. It is hard to find a chemical complex of the metal, which dissolves in the non-polar organic LS and does not deteriorate the optical property of the liquid. It is much simpler to add the metal into aqueous solutions which can be mixed into a LS with surfactants. This is one of the reasons making water based scintillators attractive.

There are several different approaches dissolving the metal in an organic liquid. One possibility is to search for a solvent with high solubility for polar compounds such as alcohols. In the scintillator of the CHOOZ experiment~\cite{CHO} the Gd salt Gd(NO)$_3$ was first dissolved in hexanol. The mixture was then diluted in other organic solvents. The case of the CHOOZ scintillator demonstrated the general difficulty in metal loaded LS production in neutrino experiments. The presence of nitrate ions in the solution deteriorated the optical properties and chemical stability of the scintillator. The degradation of the attenuation length in the final liquid limited the lifetime of the detector to about 1 year~\cite{CHO}. A solvent with high solubility for polar materials is phenethyl alcohol. With this solvent groups succeeded to dissolve a metal trifluoro-acetate at several percent loadings~\cite{PFE, PAY}. However, typical attenuation lengths at more than 5\% metal loading stayed below 1~m.

A completely different approach to dissolve metal in a LS is to use a surface active agent (surfactant) with hydrophilic (polar) as well as hydrophobic (nonpolar) chemical groups. This ``emulsive'' liquid enables polar metal compounds to be mixed into nonpolar solvents. For example, InCl$_3\cdot$4H$_2$O (Indium trichloride tetrahydrate) was mixed with a xylene based emulsive LS containing such a surfactant. With this method metal loadings of more than 5~wt.\% with an attenuation length of more than 1~m could be achieved~\cite{SUZ}.

The most promising procedure for metal loading is probably the preparation of a organometallic complex that is soluble in the LS. The most common choice is the use of metal carboxylates for this purpose. Among the carboxylic acids there are many different candidates and several of them were tested in neutrino experiments. Already Reines and Cowan were using cadmium octoate (2-ethylhexanoic acid) in their Savannah River neutrino experiment in the 1950s~\cite{RCS}. 

Several decades later a reasonably stable Gd-loaded LS using carboxylic acids was produced in the context of the Palo Verde neutrino experiment~\cite{PV1}. Again a metal 2-ethylhexanoate was used, but this time with gadolinium (Gd) instead of cadmium. Modern successfull Gd-carboxylate LS, as used in the Daya Bay~\cite{DB2014} and RENO~\cite{RENO} experiments, are using a 9C acid, 3,5,5-trimethylhexanoic acid (TMHA). A related method for metal loading is the use of naphthenic acids, which can be classified as cyclo-paraffin carboxylic acid with the typical formula C$_n$H$_{2n-2}$O$_2$~\cite{Ronzio}. The metal salts of these acids are insoluble in water, but soluble in hydrocarbons. There were also tests with phosphor-organic compounds, which were used for stabilizing purposes or to achieve solubility~\cite{Dan03, Yeh2007}. 

An alternative approach to the carboxylic acid systems is the application of beta-diketones. In general, such metal complexes are expected to be more stable than the carboxylic acid systems. The stability of the complex at temperatures above 200$^\circ$C and the rather high vapour pressures are beneficial in terms of purification processes. Such systems were studied for indium-loaded scintillators with potential use in solar neutrino experiments~\cite{In1}. The first application in a large scale neutrino detector was within the Double Chooz experiment~\cite{DC1}.

\subsection{Carboxylic acid systems}
Much of this organometallic chemistry has already been developed in the fields of separation chemistry and chemical treatment in the nuclear fuel cycle. Several organic complexing agents come to mind, much of the focus on the carboxylic acids is described in this section. Carboxylic acids are produced in bulk by the chemical industry and have lower cost and ease of chemical waste disposal, compared with other phosphorus-containing compounds. The selection of carboxylic acid for metallic extraction depends on the chemical reactivity and solubility of the metal of interest with the organic liquid. Each liquid scintillator might require different extraction strategies and complexing ligands for different metallic ions.
 
A range of carboxylic acids with alkyl chains containing from 2 to 9 carbons has been studied by early developments of the LENS solar neutrino project~\cite{Dan05}. The higher the carbon number, the lower is the acid strength and the solubility of water. Acetic acid, HAc (C2) and propionic acid, HPPI (C3), were found to have very low efficiencies for extraction of the metallic ions into the organic phase. Isobutyl acid, HIB (C4) and isovaleric acid, HIVA (C5) both have unpleasant odors and require the presence of other components as neutral organophosphorus compounds (e.g.~TBPO) to achieve high extraction efficiencies~\cite{Dan05}. Carboxylic acids containing more than 6 carbons extract the metal efficiently due to their increasing non-polarity. The rule of thumb is that the longer-carbon-chain carboxylic acids are more organic-like and thus their organometallic complexes can be dissolved in the organic liquid more easily. However, since the light yield is dominated by the weight percent of the LS, the criterion is to select the shortest possible carbon-chain acid (i.e., the lightest molecular weight carboxylic acid) that will achieve the goals of suitable optical property and chemical stability. 

The best candidate found to date for Gd or In in pseudocumene is 2-methylvaleric acid, HMVA (C6). Linear alkyl benzene, first proposed by SNO+, is the recent scintillator in favor for several neutrino experiments due to its high flash point and low cost. However, LAB has lower solubility for organometallic compounds and lower light yield than PC. The trimethyl hexanoic acid (TMHA) was found to be the most suitable complexing ligand for Gd extraction, in particular in LAB~\cite{Yeh2007, Har}. It was used in the target LS of the Double Chooz mockup as well as in the Daya Bay and RENO detectors (see section~\ref{Reactorexp}).

\subsubsection{Synthesis and Purification}
The loading of metal carboxylates into scintillator can be achieved either by liquid-liquid extraction in a two phase system of an organic LS and an aqueous phase or alternatively by solid dissolution. In the first approach, a large quantity of metal carboxylates can be solvent-extracted into the LS directly in one step within a few hours. The solid-dissolution approach needs first preparing the metal-carboxylate solids, followed by dissolving it in the organic scintillator. Although the process of solidifying the compounds involves several steps that could enhance the probability of introducing impurities during the preparation, with careful control of production procedures, both methods give similar properties of metal-doped liquid scintillator in terms of light-yield, optical transmission, and chemical stability. The synthesis of metal carboxylates, such as indium, ytterbium, and gadolinium, as neutrino- or neutron-capture targets in LS was developed through studies for solar neutrino experiments~\cite{Ra76, Ra02}. Both ytterbium and indium can be doped into PC or LAB as high as 8-10~wt.\%. The implication of gadolinium carboxylates was further focusing on the reactor antineutrino oscillation experiments, such as Palo Verde, Daya Bay, and RENO, at 0.1-0.2\%.
 
The requirement of purity for the metal-doped scintillator depends on the physics goals of the experiments. Special cares are often given to the inherent radioactivity, radon prevention, optical cleanness and chemical quantitative analysis, particularly the metal/H and C/H ratios. Most of the purification steps have to be applied before and during the synthesis of the metal carboxylates. Chemical separation schemes that would be used after the metal-doped scintillator has been synthesized are unsuitable because they would likely remove some of the metallic ions along with the other inorganic impurities. 

The developed purification techniques are divided into two categories: the removal of chemical and of radioactive impurities. The levels of non-radioactive chemical species that can adversely affect the optical property of the scintillator have to be strictly controlled. Indeed, the removal of chemical impurities not only increases the light transmission in the scintillator, but also enhances the long-term stability of the metal-doped LS. Some impurities could induce slow chemical reactions that would cause the transparency of scintillator to deteriorate gradually. 

Note that the quality and impurities of the carboxylate scintillator can vary with different vendors even at the same quoted purity level, as previously mentioned. Thus the properties of scintillator, such as its metal-loading capability, photon production yield and optical property, might change from one vendor to another and even from one production batch to another. To avoid these problems, several chemical purification steps are developed for quality assurance during the synthesis phase:

\begin{enumerate}
\item The purification of chemical ingredients in the aqueous phase, including ammonium hydroxide and ammonium carboxylate, can be done by solvent extraction with toluene mixed with 0.5\% by weight of TBPO. 
\item The low-volatility organic solvent, LAB or PC, is purified by dry column absorption (e.g.~50-cm long glass column loaded with activated neutral Al$_2$O$_3$ of 150~mesh, $\sim 58$~A particle size) or multi-staged vacuum distillation (see section 2.5.2). 
\item The high-volatility or viscous liquids, such as the carboxylic acids, can be purified by temperature-dependent, thin-film vacuum distillatory using a Teflon wiper for layer formation of inlet solvent within the distillation column and a mechanical rotary vacuum pump to keep the system pressure $<0.01$~bar.
\end{enumerate}
 
Different procedures have to be used for inorganic species such as Gd$^{3+}$ and other lanthanides, which often contain naturally occurring radioactive impurities, such as Th$^{4+}$, U$^{4+}$ and U$^{6+}$ (as UO$_2^{2+}$). Purification steps will have to be developed to reduce these radioactive species to the concentration of requirement (i.e.~$10^{-12}$~g/g in the 0.1\% Gd-LS for Daya Bay). Conventional purification methods are cation exchange (i.e. Chelex-100) and solvent extraction such as TBP, TOPO, TBPO, and TTA used in chemical processing in the nuclear industry. However, both techniques require large chemicals or exchange resins that would require special chemical handling and are not cost-effective. A new self-scavenging method utilizing different metal-hydrolysis processes by pH-control, followed by micro-filtration (2 micron) is developed to purify gadolinium from its lanthanide and actinide impurities. In this study, thorium and its decay daughters are completely removed from Gd at pH~$>6$~\cite{Yeh2010}.

\subsubsection{Characterization and Performance}
The carboxylic acids often exist in dimer forms and could be deteriorated by photo-induced oxidation in the presence of oxygen. A freshly purified batch of carboxylic acid should be prepared within 24 hours before the metal-complexing synthesis. On the other hand, after interaction with metallic ions, such optical degradation is greatly reduced due to the formation of metal-oxygen bonds. In the reactions with indium or gadolinium, the metal carboxylates show no severe degradation effect as does pure carboxylic acid. Yet, for precaution, keeping them in the dark under a nitrogen environment is highly preferred, not only to extend the shelf-life of metal-carboxylate LS, but to avoid the oxygen quenching that could reduce the light yield by 15 to 20\% for air-saturated scintillators. 

The scintillator emission spectrum is adjusted using wavelength shifters to the most sensitive region of the PMTs, typically to a wavelength range around 420 to 450~nm. The dissolution of purified metal carboxylate at few tenths of percent (i.e. Gd at 0.1~wt.\%) does not significantly affect the transmission of scintillation emission in this wavelength region. 

The development of loading metal carboxylates in either singular or binary scintillator systems is mature. The metal carboxylates can be dissolved up to several weight percent in a variety of scintillators including PC, PXE, PCH, LAB or their mixtures with n-dodecane or mineral oil. A very consistent procedure, with $>95$\% Gd chemical extraction efficiency for preparing 0.1-0.2\% Gd-loaded PC or LAB has been established for reactor antineutrino experiments~\cite{Yeh2007, DBLSP}. Ideally the nature of scintillation photon yield is a function of weight fraction of fluors in the scintillator. A scintillation cocktail of linear alkyl benzene (e.g.~Cepas) with the wavelength-shifting fluors of PPO (3~g/L) and bis-MSB (15~mg/L) gives $9200-9500$~photons/MeV depending on the vendor and method of preparation. Carboxylate is a self-quencher that could affect the FRET between fluors in the LS and should be controlled at minimal level, just enough for complexing metallic ions in the LS. For instance, 0.3~wt.\% of carboxylic acid in 0.1~wt.\% Gd-doped LS gives about 90\% light yield relative to the metal-free LS version. 

Based on LENS R\&D activities on Yb, In and Gd-LS by an INFN(Italy)/INR(Russia) group~\cite{Dan03, Dan05, Dan07}, a Gd-LS was produced for supernovae neutrino detection in the LVD experiment at LNGS, Italy~\cite{LVD}. Two modules of the LVD detector, each of them containing 1.2~tons of LS inside a 1.5~m$^3$ stainless steel tank, were loaded with about 1~g/l Gd to increase the signal to noise ratio for $\bar\nu_e$ detection. The unloaded LS in the LVD experiment is a mixture of aliphatic and aromatic (8 or 16\%) hydrocarbons known as White Spirit or Ragia Minerale. The light attenuation length of this liquid at 425 nm is well above 10~m. The Gd compound used was prepared as solid Gd-2-Methylvalerate salt following the procedure developed in~\cite{Dan07}. A 15~kg batch could be produced in one month using standard equipment of a chemical laboratory. The salt was dissolved in a highly concentrated solution and subsequently diluted in the tank. The stability and performance of the 2.4~tons of Gd-LS were tested. Attenuation length after Gd-loading was still above 10~m in one of the tank. For the other tank a reduced transmittance with an attenuation length of about 6~m was observed directly after loading. From periodical measurements in an UV/Vis photospectrometer over three years stable behaviour of the attenuation length was observed in both tanks. A degradation of the absorbance of more than 1\%/year could be excluded from the measurements. 

\subsection{Beta-diketone systems}
Whereas the two oxygen atoms of the carboxylic acid molecule are bound two the same carbon atom, the $\beta$-diketone (BDK) molecule contains two ketone groups, which are separated by a methyl group. After removal of a proton the negative BDK anion can bind to the metal in a chelate ring structure by resonance stabilization~\cite{Har}.

Metal BDK systems were used for several different elements in neutrino physics. The first application of this type was in the context of neutrino mass measurements using the calometric measurement of low energy electron capture decays of the rare earth isotope $^{163}$Ho in a gas proportional counter~\cite{Hol}. The beneficial properties of the organometallic BDK complex as thermal stability and volatility were already pointed out in the 1960s when it was shown that those complexes can be used to separate rare earth elements by gas chromatography of the BDK complexes~\cite{Eis65}. Beyond neutrino physics such complexes can be e.g.~used as NMR shift reagents~\cite{Sie73} or as fuel antiknock additives~\cite{Eis74}.

The first proposal to use these metal complexes for neutrino search in a LS detector was in the context of the LENS project~\cite{In1}. The focus of the project was to study solar neutrinos using an indium loaded LS. Here, the advantages of hydrolytic and oxidative stability and the characteristic solubility of metal BDK molecules in non polar organic solvents were used. Metal loadings of more than 5~wt.\% could be achieved with attenuation lengths in the meter-scale. Transparency and radiopurity of the complex was significantly improved in a sublimation process of the powder material. One of the main challenges in such systems is to minimize the quenching of the light yield. Similar to the metal carboxylate system, the energy transfer from excited solvent molecules to the primary fluor is in direct competition with the energy transfer to the non-fluorescent metal complex.

The first use of this BDK-scintillator systems in full scale neutrino experiments was within the Double Chooz~\cite{DC1} and Nucifer~\cite{Nuc} reactor experiments. Here, the metal of choice is Gd, which has the highest cross section of all elements for thermal neutron capture. In this way the detection characteristics of the neutron created in the neutrino reaction is improved significantly. The metal loading in such applications can be more than one order of magnitude lower (0.1~\%) than for the case of an In-loaded solar neutrino detector. The scintillators in Double Chooz and Nucifer showed chemically and optically stable behaviour during several years of data taking. 

An organic LS containing a zirconium complex based on the BDK technology has been developed for a new neutrinoless double $\beta$-decay experiment~\cite{ZrBDK}. A solubility of more than 10~wt.\% was found with anisole as a solvent. To reduce light quenching a $\beta$-keto ester complex was synthesized introducing –OC$_3$H$_7$ or –OC$_2$H$_5$ substituent groups in the BDK ligand. In this way the absorption peak was shifted to shorter wavelengths away from the solvent emission peak. Since the shift of the absorption peak depends on the polarity of the solvent a low polarity solvent should be used.

\subsubsection{Characteristics and variations}
The remarkable stability of BDK metal complexes is mainly due to the strong chemical bond of the $\beta$-diketone anions to the centrally chelated metal ion. In this way, the metal gets trapped in a stable organic structure and creates a compound that can be dissolved in organic LS. The simplest form of a $\beta$-diketonate is acetylacetone (C$_5$H$_8$O$_2$, 2,4-pentanedione, commonly Hacac). The In form of this BDK version (In(acac)$_3$) was successfully tested in the LENS Low Background Facility at the Gran Sasso underground laboratory in Italy~\cite{In2}. The Hacac was mainly chosen as ligand to the In metal because of the low molecular weight. The lower the weight of the non-fluorescent metal compound, the higher the weight of the components participating in light production in the LS.

In many applications the metal to be loaded to the scintillator is a rare earth as gadolinium (reactor neutrinos), neodynium ($\beta\beta$-decays) or ytterbium (solar neutrinos). Other than In(acac)$_3$, trivalent rare earth acetylacetones tend to form hydrates (M(acac)$_3\cdot$nH$_2$O). The literature indicates that basically only two hydrates are formed: the monohydrate (n=1) and the trihydrate (n=3)~\cite{Pope}. The trihydrates is the most common form after synthesis. From that molecule the monohydrates can be produced by drying methods or by crystallization from a solution of the trihydrate in e.g.~ethyl alcohol or acetone. The hydrates of the rare earth elements with lower atomic number are more stable. The larger diameter of the latter probably enable them to accomodate more easily the additional water molecules.

The removal of the last water molecule in the monohydrates might be achieved under vacuum at elevated temperature of about 60$^{\circ}$C~\cite{Liss}. However, the material could decompose in this step and partially be converted to M(acac)$_2$OH which is reported to be stable for Gd~\cite{Prz}. To avoid such complications and to improve the solubility in organic solvents the Hacac ligand can be replaced by bigger BDK molecules as Hthd (C$_{11}$H$_{20}$O$_2$, 2,2,6,6-Tetramethylheptane-3,5-dione). This more stable form was chosen e.g.~for the reactor neutrino experiments Double Chooz~\cite{DC1} and Nucifer~\cite{Nuc}. 

The solubility of the BDK metal complexes in organic solvents might depend on the crystal structure of the powder material. In general, correlation between high vapor pressure and high solubility can be found~\cite{Meh}. For Gd(thd)$_3$ high solubility is found in ethanol and very high solubility in tetrahydrofurane (THF). The latter can be used to break up the crystal structure of the powder and its presence in the scintillator has a stabilizing effect on the liquid. One or more THF molecules can stick to the complex when the coordination number of the central metal atom is above six. In this way it protects the complex from attacks of other molecules as water, which could lead to decomposition. In most complexes volatility, thermal stability and solubility are increased when fluorine substituents are present in the ligand shell~\cite{Meh}. An example for such a BDK molecule is heptafluorodimethyloctanedione (C$_{10}$H$_{11}$F$_7$O$_2$, Hfod), which on the other hand is typically limited in terms of transparency.  

\subsubsection{Synthesis}
In the synthesis of the metal BDK molecules the starting material is typically metal chloride, oxide or nitrate. Below the basic synthesis steps for two different metal BDK systems, In(acac)$_3$ and Gd(thd)$_3$ are described. For the production of In(acac)$_3$ as described in \cite{In1}, InCl$_3$ is first dissolved in water at a pH around 2 and Hacac is added to the solution. Then the indium material is transferred into an organic phase by solvent extraction. For this an organic solvent such as 1,2-dichlorobenzene, which has high solubility for In(acac)$_3$, is added and the solution is neutralized using a base (NaOH solution). One of the main challenges in the synthesis is to avoid hydrolysis as much as possible and the removal of hydrolysis products. After approximately one day the organic phase containing the indium is separated from the aqueous layer. The solvent is removed under vacuum at elevated temperature resulting in the formation of white In(acac)$_3$ crystals. The crystals can be purified by washing them in 2-propanol in which the solubility of In(acac)$_3$ is negligible. The final material is obtained after filtering and drying and can be dissolved directly in the LS, typically up to indium concentrations of around 1~wt.\% and above. 

A way to synthesize trivalent rare earth complexes as Gd(thd)$_3$ with yields of more than 90\% is described in~\cite{Eis65}. The H(thd) is first dissolved in ethanol and connected to a vacuum system. To reduce impurities in the H(thd) it might be distilled before usage. Then a solution of NaOH dissolved in a 50\% ethanol/water mixture is added while stirring. The metal nitrate, also dissolved in 50\% ethanol/water, is added to the reaction flask. The solution in the flask is stirred under vacuum and its volume reduced by evaporation of the ethanol. The metal complex precipitates after the addition of pure water. The crystals get isolated by vacuum filtration and are dried after washing with water and/or a water/ethanol mixture. As a final step the material should be sublimed to reach highest purity of the crystals.

\subsubsection{Sublimation and chemical purity}
The stable structure of metal BDK complexes allows heating of the compound without decomposition and in combination with the typically high vapor pressures purification steps by sublimation can be applied. For the sublimation the sample material is placed into a glass tube. The dimensions of the tube depend on the sublimation rate to be achieved. A vacuum is applied and the metal BDK complex is heated, typically to around 200$^{\circ}$C, using a heating mantle. The vacuum is drawn at the exit of the tube while a slight flow of N$_2$ or argon carrier gas enters from the upstream side. The vapor condenses downstream in an unheated region and pure crystals start to grow~\cite{In1}. With this technique purification rates up to 1~kg/day could be achieved~\cite{LSP}.

In the sublimation step the complex gets separated from impurities as partially hydrolyzed components or BDK complexes with other central metal ions due to different sublimation parameters. Furthermore, sublimation can improve significantly the radiopurity of the sample. In some samples the potassium contamination was found to be more than two orders of magnitudes lower for the sublimed compared to unsublimed material~\cite{LSP}. The absorbance of the powder in the wavelength region of interest can be reduced in the sublimation step by a factor 2 and more depending on the purity of the starting material. Molar extinction coefficients well below 0.01~liters/(mol$\cdot$cm) above 420~nm were reached~\cite{In1, LSP}. At such purity levels the transparency of the LS is typically dominated by the primary fluor and several meters of attenuation length can be achieved even with metal loadings of few per cent. The influence of the metal BDK on the transparency of the LS is usually weak and less critical than the impact on the light yield caused by quenching effects. 

\subsubsection{Quenching}
The addition of a metal complex to the LS opens an additional energy transfer path from the solvent molecules to the typically non-fluorescent metal molecule. This transfer is in competition with the transfer to the primary fluor, which is relevant for scintillation light production. The energy transfer to the metal BDK molecule can be rather effective, since the absorption spectra are in similar wavelength regions than the one of the primary fluor, around 300~nm~\cite{Bu07}. For the same molar concentration of primary fluor and metal complex the energy is sometimes even more likely transferred to the metal compound than to the fluor~\cite{LYA}. This means for reasonable light yields ($>50$~\% of the unloaded scintillator) the fluor concentrations should be about as high as the metal concentration. However, at high metal and fluor loadings self-quenching and self-absorption of the primary fluor become significant.

\subsection{Quantum dots}
In recent years new methods for the production of metal loaded LS in neutrino physics were studied involving semiconducting nanocrystals, which are known as ``quantum dots''~\cite{Wins}. The optical and electrical properties of the quantum dots are directly proportional to their size, which is typically in the order of few nanometers. Smaller dots absorb and emit photons at shorter wavelengths. The emission band consists of a narrow resonance around the characteristic wavelength of the dot. Since the dot size can be controlled to high precision in the synthesis, the absorption and re-emission spectrum of the dots can be tuned and optimized for a respective application. In particular, the emission spectrum can be matched to the region of maximal detection efficiency in the photosensor or it is optimized for highest energy transfer to the secondary fluor. The organic shell of the quantum dots promotes loading into organic solvents. In some synthesis methods the quantum dots are already delivered in colloidal suspension with the aromatic solvent toluene at concentrations of several grams per liter.

The most commonly used quantum dot cores are binary alloys such as CdS, CdSe, CdTe, and ZnS. Alternatively, there are also phosphor-based rare-earth dots. Therefore, quantum dots provide a method to dope scintillator with various metals and rare-earth elements. For a cadmium (Cd) based LS there are even two different applications in the field of neutrino physics. Since the isotope $^{113}$Cd (natural abundance 12.2~\%) has a very high thermal neutron capture cross-section with a gamma cascade of about 9 MeV, it is a good candidate for antineutrino measurements using inverse beta decay. Moreover, Cd is also interesting because it includes two isotopes which are $\beta\beta$-decay candidates ($^{106}$Cd and $^{116}$Cd). Also Se, Te, and Zn, which are present in common quantum dot cores, have $\beta\beta$-decay candidates.

For typical solutions of quantum dots in toluene at concentrations around 1~g/l, attenuation lengths of less than 1~m were found. However, after filtering the attenuation length can increase by a factor 10 and values in the few meter scale are observed in the region of interest~\cite{QDAb}. This is close to the requirements of large scale neutrino detectors. The concentration of quantum dots in the liquid was found to be stable before and after the filtering step. Stability tests indicate that larger particles are formed by aggregation in the concentrated solutions over long time scales. This could explain the fact that filtering improves the attenuation length as well as the observed transparency degradation after several weeks~\cite{QDAb}. The quantum yields of standard quantum dots are in the range from 30--50~\%~\cite{QDAb}, which is a factor two below the values of typical wavelength shifters as PPO or bis-MSB~\cite{Ber, QY, QYDB}. However, there is room for optimization on this parameter by improvements in the process of organic shell growing. 

In the context of quantum dot loaded scintillators there is also a proposal to tune the scintillator emission spectrum in a way that allows to separate scintillation from Cerenkov light~\cite{Wi14}. Ideally there would be sharp narrow emission from the quantum dots (scintillation) at shorter wavelengths separated from the Cerenkov contribution extending to rather long wavelengths. With an optimized selection of the photosensor efficiency the Cerenkov to scintillation light ratio could be tuned. The advantage of the Cerenkov light is that it contains directionality information in contrast to the isotropic emission of scintillation light. If the Cerenkov light can be isolated from scintillation light, the additional directionality information might provide powerful background rejection techniques, e.g.~for neutrinoless $\beta\beta$-decay searches.

\subsection{Water based liquid scintillators}
A new type of scintillation medium for large liquid detectors, in which scintillating organic molecules and water are co-mixed using surfactants has been developed. These water based liquid scintillator (WbLS) systems~\cite{WLS11, WLS15a} are flexible, with the ability to form scintillation solutions over a wide range of compositions that can be tuned to meet detector needs ranging from almost pure water to almost pure organic for different nuclear and particle physics experiments. While pure water detectors have been employed in large neutrino detectors in the past, they are primarily sensitive to high energy interactions creating particles above the Cerenkov threshold.  

WbLS is a novel cost-effective detection medium that will enhance future massive Cerenkov detectors with the unique capability of scintillation detection to be sensitive to low-energy physics below Cerenkov threshold and being considered in uses for numerous future frontier experiments. A deployment of such a scintillating water detector with careful controls of low-energy radioactive background could improve the current proton-decay sensitivity ($\sim 10^{33}$~y) by an order of magnitude~\cite{WLS11} and have additional potential for neutrino and rare processes in next-generation experiments. 

\subsubsection{Synthesis} 
Organic liquid scintillator dissolved at sufficient concentration in a water medium is essential to the success of water-based liquid scintillator. Chemically organic solvents are immiscible in water mainly due to the differences in polarities. It will entail a surfactant that contains lypophilic and hydrophilic groups to emulsify the organic liquid scintillator into the water solvent. Engineering of a complexing medium to stabilize the lypophilic and hydrophilic molecules in a water medium with appropriate optical transmission and long-term stability is essential. The solubility of a series of conventional organic liquid scintillators, such as PC, PXE, PCH, LAB, commercially capable in large quantity, have been systematically studied in water. Their loading capabilities, optical properties, purification methods, proton densities and commercial availabilities are well understood. A suitable ampliphilic surfactant or mixtures of two surfactants that can reduce the liquid-liquid interfacial tension between organic solvent and water, and thus blend the LS into the bulk water molecules, are necessary to stabilize the WbLS solution. The degree of tension reduction is a function of surfactant concentration, which could also affect the optical and stability properties of the medium.

\subsubsection{Preparation and Performance}
It should be noted that loading of a wavelength shifter into water is different from the water-based liquid scintillator. The former only improves the Cerenkov counting efficiency; but has no capability of capturing incident UV photons induced by ionization radiation. Linear alkyl benzene currently used in several neutrino experiments (Daya Bay, SNO+, Reno, JUNO) is the scintillator solvent of favor for WbLS production. LAB has a long alkyl chain, 10-13 carbon atoms, attached to the benzene ring (where the unsaturated $\pi$-electrons are essential for emission of ultra-violet light due to incident ionizing radiation). The worldwide industrial uses of LAB (millions of tons per year) are for lubricants and organic solvents in application of manufacturing detergents, in which a sulfonic-acid (HO-S-O-OH) (LAS) or amine derivative (PRS) of alkyl benzene is produced. Briefly, the chemical basis of linear alkylbenzene sulfonate derivatives is that the alkyl chain and benzene group act as a solvent to extract oils and grease (lypophilic), while the attached sulfonate or amine group is soluble in water (hydrophilic). The unique lypophilic and hydrophilic properties, and associated benzene $\pi$-electron ring, acting as surfactant and liquid scintillator simultaneously enable either PRS or LAS to be a promising anionic surfactant for preparing a water-based liquid scintillator (or organo-scintillator loaded water detector). Similar results can also be achieved by nonionic surfactants that have a hydrophilic polyethylene oxide chain (on average of 9.5 ethylene oxide units) attached to an aromatic hydrocarbon lipophilic or hydrophobic group (NP-4,-40, TX-100,114,405, etc.). The production quantity and cost of these surfactants are very attractive for large-scale neutrino detectors.

New water-based liquid scintillator R\&D based on these surfactants was initiated by the BNL neutrino and nuclear chemistry group in 2010. A series of WbLS samples loaded with different percentange of organic scintillator in water in use for different physics applications have been successfully synthesized and stable for years since preparation. A $>$kt scintillation water detector~\cite{Theia} capable of Cerenkov and scintillation detections needs long attenuation length at tens of meters in the PMT sensible region. In the example of lightly organic doped (1~wt.\%) water, its absorption length is superior to conventional pure LS detectors; however the main challenge of optical transmission is the scattering effect introduced from the inhomogeneous charge distribution between the water and organic interfaces. A mixture of non-ionic and ionic surfactants could improve the scattering significantly. The WbLS principal could also be applied to loading of metallic ions directly in aqueous solution to provide sensitivity for a wide range of particle interactions with good light yield and high transparency in large detectors. Such an approach of extracting the hydrophilic elements (e.g. B, Te, and Li) into organic scintillators cannot be achieved using conventional organometallic technology using carboxylate or phosphate ligands. Water-based detectors could also serve as the near detector for long-baseline neutrino beam monitoring or be used for detection of diffuse neutrino flux from distant past supernovae. 

Potential WbLS contributions to proposed LS experiments are
\begin{itemize}
\item SNO+ \cite{SNO+}: This experiment plans to use a Te-doped LS to determine the neutrino mass and Majorana or Dirac nature of the neutrino via neutrinoless double beta decay. Formulation of a high Te-loaded scintillator (3~wt.\%, phase-II) to increase the sensitivity of a ton-scale $0\nu\beta\beta$ detector using either water-based or diol-complexing loadings, 
\item LUX-ZEPLIN (LZ) \cite{LZ}: In the context of the upcoming LZ dark-matter detector a Gd-LS with low U and Th background ($<$~ppt) is developed serving as a veto system for dark matter search. A bench-top sample at 5~wt.\% is currently under evaluation, 
\item PROSPECT \cite{Pros}: Prospect will use a $^6$Li-doped LS (0.1~wt.\%) for the detection of reactor antineutrinos at short baselines ($<10$~m). The $^6$Li-LS with long stability, reasonable light-yield and attenuation length, and high background rejection capability via pulse shape discrimination (PSD) is ready for ton-scale production, 
\item T2K near detector \cite{T2K}: A water-based LS with scintillation and Cerenkov features for neutrino beam physics and improved control of neutrino-water cross section uncertainties for neutrino oscillation measurements is tested, 
\item JSNS2 \cite{Hara}: A Gd-doped WbLS detector deployed at J-PARC Spallation Neutron Source to search for an eV$^2$-scale sterile neutrino, using both neutrinos and antineutrinos as well as both electron-flavor-appearance and disappearance, and muon-flavor-disappearance channels is under evaluation, 
\item THEIA \cite{Theia}: Within the project water-based LS at various wt.\% loadings are under development for multi-physics topics in nuclear, high-energy, and astrophysics, ranging from a next-generation neutrinoless double beta decay search capable of covering the inverted hierarchy region of phase space, to supernova neutrino detection, nucleon decay searches, and measurement of the neutrino mass hierarchy and CP violating phase, and
\item Medical Imaging: A scintillating water phantom for quality assurance of ion-beam therapy (Patent) or a photon detector (loaded with high Z element) for TOF-PET.  The radiation hardness has been studied that a scintillating phantom (5\% loading) used for proton therapy treatment plan verification would exhibit a systematic light yield reduction of approximately 0.1\% after a year of operation, which is highly feasible in such applications~\cite{WLS15b}.
\end{itemize}
The detector requirements for each experiment are summarized in Table~\ref{water}.

\begin{table}[h]
\caption[waterprojects]{Scintillator detector design for projected experiments. The percentages given for the WbLS correspond to the amount of organic scintillator in the liquid. \label{water}}
\begin{center}
\begin{tabular}{ll}
Experiment & Detector requirement if using WbLS\\
\hline
SNO+ & WbLS (70\%+) doped with 3\%$^{130}$Te ($0\nu\beta\beta$ isotope)\\ 
LZ & Pure LS doped with 0.1\%Gd with U/Th $<$~ppt\\
PROSPECT & WbLS (80\%+) doped with 0.1\% $^6$Li with high PSD\\
T2K & WbLS (10\%)\\
THEIA & WbLS (1\%)\\
QA phantom for ion-beam therapy & WbLS ($1-5$\%)\\ 
TOF-PET calorimetry & WbLS (10\%) loaded with high Z elements\\
\hline
\end{tabular}
\end{center}
\end{table}

\subsubsection{One-ton WbLS Demonstrator}
The principal of a mass-producible, cost-effective water-based liquid scintillator using the derivatives of linear-alkly-benzene- or hydrophilic polyethylene oxide-based surfactant has been proven. For the success of WbLS application to large-scale physics experiments, several key characteristics of adequate photon production, long attenuation length, quenching effect and timing property, in addition to chemical stability, need to be demonstrated. The preliminary data show that a 1~wt.\% WbLS could emit 100 optical photons per MeV with ~30~m attenuation length (dominated by scattering) at 450~nm. Nevertheless, on-line purification schemes for the organic-water mixing system need to be developed. A one-ton prototype is currently installed at BNL and studies with calibration radioactive sources, cosmic ray muons, and/or high intensity proton beam (100~MeV to 1~GeV) are planned. The goal is to examine the light propagation phenomenon including scattering and quenching effects, direct Cerenkov separation from scintillation and solution stability under various environments. Loading organometallic ions into WbLS to enhance the particle-capture signal is well-demonstrated by SNO+ (Te) and PROSPECT ($^6$Li) experiments and ton-scale production of metal-loaded WbLS is under development. Preliminarily loading of gadolinium into the developed WbLS can be achieved at $>0.5$~wt.\% in bench-top studies; few loading parameters to further improve the scattering length and light-yield performances still need to be fine-tuned. Although each metal ion favors specific surfactant for loading in scintillator, the principal and procedure are well established.

\newpage
\section{Applications in neutrino experiments}
\subsection{Reactor neutrinos}
\label{Reactorexp}
Metal loaded liquid scintillators were used in most of the neutrino experiments at nuclear reactors. Nuclear reactors are a pure and strong source of electron antineutrinos. The fission fragments of the uranium and plutonium isotopes in a nuclear reactor almost exclusively undergo $\beta^-$-decay. For each fission about six electron antineutrinos are produced corresponding to more than $10^{20}$/s isotropically emitted neutrinos per GW of thermal power. For detection the reaction on free protons of hydrogen (H) atoms in scintillator molecules can be used: $\bar\nu_e+H^+\rightarrow n+e^+$. The energy threshold for this reaction is at 1.8~MeV. The emission of a positron (up to $\sim10$~MeV energy) and a neutron create a coincidence signal in the liquid. A prompt event from the positron carries the energy information of the incident antineutrino. The neutron thermalizes in the liquid within few centimeters and is captured on a nearby nucleus. In case of a standard liquid scintillator without any metal loading the neutron is captured mainly on H on a time scale of about 200~$\mu$s. This neutron capture on H with the emission of a 2.2~MeV photon marks the delayed event in the coincidence signature. 

The coincidence signal allows for efficient background reduction. Nevertheless, due to the rather low energy of the gamma in the energy region of natural radioactivity (up to about 3~MeV) the background rate with an unloaded LS would be still too high for most of the reactor neutrino experiments. Therefore elements with high cross sections for thermal neutron captures as Cd or Gd are added to the liquid. This improves the signal to background ratio in two ways. First, the coincidence time window can be shorter. Second, the gamma energy after neutron capture is higher for those elements. Both effects effectively reduce accidental background due to random coincidences. Nowadays most experiments use Gd, since the isotopes $^{155}$Gd and $^{157}$Gd of this rare earth element have the highest cross sections for thermal neutron capture among all stable natural isotopes. The total energy of the emitted gammas after neutron capture is about 8~MeV. This is well above the typical energies of natural radioactivity from K, U or Th and their daughter nuclei. Below the performance of metal loaded LS in selected reactor experiments is described, which were pioneering the field or were of major relevance for scintillator research. Basic properties of the scintillators used in those experiments are listed in Table~\ref{ReactorLS} 

\begin{table}[h]
\caption[Reactor targets]{Selection of metal loaded scintillators used in reactor neutrino experiments. The attenuation length is given in the region 430--440~nm and the light yield in \% of anthracene. \label{ReactorLS}}
\begin{center}
\begin{tabular}{lcccccc}
Experiment & solvent & metal & system & ligand & att.~[m] & light yield \\
\hline
Sav.~River & TEB/oil & Cd & carbox. & 2-ethylhex.(C$_8$) & 2~m & \\
Bugey & PC & $^6$Li & carbox. & salicylate & 2.6~m & 31\\
CHOOZ & IBP/paraffin/ & Gd & nitrate & (NO$_3$)$_3$ & 4~m & 35\\
      & hexanol &  &  &  &  & \\
Palo Verde & PC/oil & Gd & carbox. & 2-ethylhex.(C$_8$) & 10~m & 56\\
Double Chooz & PXE/dodec. & Gd & $\beta$-diketone & THD & 10~m & 38\\ 
Daya Bay & LAB & Gd & carbox. & TMHA (C$_9$) & 15~m & 50\\
RENO & LAB & Gd & carbox. & TMHA (C$_9$) & 10~m & 48\\
\hline
\end{tabular}
\end{center}
\end{table}

\subsubsection{First neutrino detectors}
Already in the early work of Reines and Cowan on first neutrino detectors, metal loaded LS was considered as key technology due to the high number of proton targets and good detection efficiency. Their first experiment searching for neutrinos was located at the Hanford Site~\cite{R53}. Later the experiment moved to the Savannah River Plant in South Carolina, since the new location provided better shielding against cosmic rays. The Hanford detector of the Reines group used toluene as scintillator solvent basis, terphenyl as primary fluor and $\alpha$-naphthyl-phenyl-oxazole as secondary WLS. The metal was loaded in the form of cadmium propionate (Cd(C$_3$H$_5$0$_2$)$_2$). The light transmission of the liquid was about 1~m. The solution deteriorated after about 1~year with the formation of an insoluble crystalline precipitate.~\cite{RCS}

Still in Hanford, Reines et al.~tested alternative components, in particular due to toxicity of toluene and Cd propionate. They searched for a more robust liquid, with optical properties less sensitive to LS handling. As solvent there were tests with mineral oil and triethylbenzene (TEB). With TEB high light yields were observed and attenuation lengths of 10~m were recorded before metal loading. The wavelength shifter combination was modified to terphenyl as primary and POPOP as secondary WLS. The carboxylate was also replaced in the scintillator version used for the Savannah River detector. Cadmium octoate (Cd(C$_8$H$_{15}$0$_2$)$_2$), the cadmium salt of 2-ethylhexanoic acid, had shown reasonable performance. At Cd:H ratios of 0.0015 attenuation lengths of 2~m and more were found with better stability compared to the Cd propionate~\cite{RCS}. This was sufficient for the needs and dimensions of the Savannah River detector at that time. 

\subsubsection{Bugey}
About 40 years after the first neutrino detection by Reines et al., a slightly different approach was followed by the Bugey 3 experiment~\cite{Bug3}. This experiment was designed to measure oscillations of reactor neutrinos using 3 identical detection modules. While most experiments use the gamma signal from Gd after neutron capture, a LS loaded with $^6$Li was used in Bugey 3. An advantage of using Li is the good confinement in space and time of the thermal neutron capture reaction $n+^6Li\rightarrow \alpha +^3H$ with kinetic energy of the $\alpha$-triton pair of 4.8~MeV allowing efficient background reduction. In addition, there are powerful pulse shape discrimination (PSD) capabilities, since the alpha and triton are strongly ionizing particles with different time profiles of the scintillation signal compared to electrons or positrons. The detector efficiency in a Li-doped LS can be kept high, even close to the walls of the detector volume. 

The Bugey 3 detector had a modular design. Rectangular LS cells (8x8x85~cm$^3$) were equipped with a photomultiplier at both ends. The combination of 98 cells formed together a detector module containing in total 600~liters of LS. One of those modules was placed at a distance of 15~m from the reactor, two clones at a longer distance of 40~m. The Bugey 3 LS was based on pseudocumene (PC) as solvent doped with 0.15~wt.\% of $^6$Li. The advantage of using PC, beyond the high light yield and associated energy resolution, is the good PSD capability of such a LS. The attenuation length was measured to be $(2.6\pm0.3)$~m at 440~nm wavelength with a light yield of $(31\pm5)$\% of anthracene~\cite{Bug3a}. Still, the Bugey scintillator showed instabilities, in particular degradation of the attenuation length. The size of the effect was different in the individual cells. The number of photoelectrons observed for a fixed energy deposition dropped by up to 4\% per month~\cite{Bug3}. Nevertheless, Bugey 3 succeeded to provide high statistics neutrino energy spectra close to nuclear reactors~\cite{Bug3b}. 

\subsubsection{CHOOZ}
Few years after Bugey, the CHOOZ~\cite{CHO} experiment tried to measure neutrino oscillations using longer baselines. The reactor neutrino experiments before CHOOZ measured the neutrino flux at distances of about 100~m and below, the CHOOZ experimental site is 1.05~km from the two reactor cores. Therefore a larger detector size was required to obtain reasonable event rates. The target mass of the CHOOZ LS was about 5~tons inside a single acrylic vessel. In such a big LS detector the pathlength of the scintillation light through the detector liquids is increasing, setting stronger requirements on the optical clarity of the liquid. 

The CHOOZ LS was doped with 0.9~wt.\% of Gd in the form of a nitrate. The salt Gd(NO$_3$)$_3$ was dissolved in hexanol. This solution was then diluted in a paraffinic liquid (Norpar-15) and isopropylbiphenyl (IPB) with p-terphenyl and bis-MSB as fluors. At the start of the experiment the attenuation length was about 4~m and the light yield 35\% of anthracene~\cite{CHO}. The scintillator transparency showed significant degradation over time, limiting the livetime of the CHOOZ detector to about one year. Oxidation by the nitrate ion was suspected to be the main reason for the observed LS degradation~\cite{CHO}. Nevertheless, with the data available it was possible to set the most stringent upper limit on the neutrino mixing angle $\theta_{13}$ for about one decade.

\subsubsection{Palo Verde}
At the time of CHOOZ another neutrino experiment at a distance of about 1~km from the reactor was constructed at the Palo Verde Nuclear Generating Station, back then the largest nuclear power plant in the US~\cite{PV1}. The antineutrino spectrum was measured using a segmented Gd-loaded LS detector. The detector consisted of a 6x11 array of acrylic cells dimensioned at 900x13x25~cm$^3$ and was filled with a total of 11.3~tons of LS. As with the CHOOZ experiment, Palo Verde did not find any hint for neutrino oscillations and found a similar upper limit on $\theta_{13}$~\cite{PV2}.

The final Gd-loaded LS used in the Palo Verde experiment fulfilled the needs of the experiment. The scintillator development however reflected the difficulties and challenges in the production of metal loaded LS. For material compatibility reasons with the acrylic cells, the LS basis PC was diluted with mineral oil (60~Vol.\%). In Palo Verde the metal oxide Gd$_2$O$_3$ was converted to the carboxylic acid 2-ethylhexanoate~\cite{Pie} and then loaded to the scintillator at a Gd-concentration of 0.1~wt.\%. 

A first batch of fully blended scintillator did not even remain stable during transport in 200~liter steel drums. Therefore, it was decided to mix a PC based concentrate on-site directly with mineral oil and additional PC. The concentrate itself was produced by Bicron company (BC-521C). Special care had to be taken during mixing, since rapid addition of the oil led to irreversible precipitation of the Gd compound. Any disturbance or agitation of the liquid was a risk for the scintillator stability. Even nitrogen bubbling for oxygen removal had to be avoided, not to destabilize the liquid~\cite{Pie}.

Whereas the light yield was found to be rather robust, the main challenge in the scintillator production was to achieve long attenuation lengths and keep them over long time periods. After several steps of optimization in the scintillator production procedure attenuation lengths of more than 10~m could be achieved. Although some aging effects were still observed, stability of the liquids transparency was acceptable for the needs of the Palo Verde experiment. The light yield was measured to be 56\% of anthracene.~\cite{Pie}

\subsubsection{Double Chooz}
The search for the last unknown mixing angle in the three flavor paradigm of neutrino oscillations triggered the construction of three additional reactor neutrino experiments at the 1~km baseline about 10 years after CHOOZ and Palo Verde. This new generation of experiments has additional near detectors at shorter baselines, which measure the neutrino flux before oscillation takes place. In this way, the systematic uncertainties from flux prediction and neutrino detection can be effectively reduced allowing to measure the oscillation effect. The first indication for disappearance of reactor neutrinos due to a nonzero value of $\theta_{13}$ was provided by the Double Chooz experiment~\cite{DC1}. 

For the Far Detector in Double Chooz, the site of the CHOOZ experiment could be reused. The detector design was improved in several ways. Among the improvements there is the larger target size (about factor 2) and the inclusion of a non-scintillating Buffer liquid as shielding, protecting the sensitive detector volumes from external radioactivity~\cite{DC1}. As solvent for the 10~m$^3$ of target liquid in each detector a mixture of 20~Vol.\% o-PXE and 80~Vol.\% n-dodecane were used~\cite{LSP}. A n-alkane was chosen instead of mineral oil to have a well defined chemical composition. In this way the H ratio can be calculated to a high precision. Furthermore n-dodecane has the advantages that it is highly transparent, the radiopurity had already been proven by the KamLAND experiment, it is compatible with the acrylic vessel and provides a large number of target protons. 

In Double Chooz for the first time a metal loaded scintillator based on the $\beta$-diketone chemistry was used on a large scale. The liquid was loaded with 1~g/l (0.123~wt.\%) of Gd which reduces the mean neutron capture time of the undoped liquid by a factor of more than 6 to less than 30$\mu$s. Stable transparency of more than 5~m above 430~nm was observed~\cite{LSP} during several years of data collection~\cite{DC3}. Special care was taken to keep the H$_2$O level below 50~ppm. The target light yield was measured to be 58\% of a pure o-PXE based scintillator with the same fluor combination of PPO and bis-MSB~\cite{LSP}. The radiopurity of the liquid was found to be well within specifications.   

\subsubsection{Daya Bay}
A similar detector design as in Double Chooz was implemented in the Daya Bay experiment~\cite{DB2012}. This experiment provides the most precise measurement of the neutrino mixing angle $\theta_{13}$ so far with sin$^2(2\theta_{13})=0.084\pm0.005$~\cite{DB2015}. A rather large number of reactors and neutrino detectors are involved. The Daya Bay nuclear power complex in China consists of six cores grouped in three pairs with maximal distances from each other of about 1~km. Neutrinos are measured in three experimental halls, two of them near the reactor cores each housing two identical detectors with 2x20~tons of target liquid. In the experimental hall at the far position close to the baseline of maximal oscillation, another four identical detectors are installed. This configuration provides very high statistics compared to other reactor neutrino experiments. 

The liquid scintillator in Daya Bay is based on LAB as solvent and uses the carboxylic acid 3,5,5-trimethylhexanoic acid (TMHA) for the Gd-loading (0.1~wt.\%)~\cite{DBLSP}. After thoroughly testing and prototyping of the liquid almost 200 tons of Gd scintillator were produced within 6 months. Solid Gd-TMHA prepared from GdCl$_3$ was dissolved directly into the LAB. The attenuation length as measured in the storage tanks was about 15~m at 430~nm, a typical wavelength for scintillator emission. The photoelectrons per MeV reduced by about 1.3\% per year due to a slight deterioation of the attenuation length~\cite{DBLSP}. A high light yield for a metal loaded LAB based LS of about 50\% of anthracene was observed.

\subsubsection{RENO}
The third reactor neutrino experiment measuring the neutrino mixing angle $\theta_{13}$ is the RENO experiment~\cite{RENO} in South Korea. As neutrino source six pressurized water reactors are positioned along a 1.3~km long straight line. Two identical antineutrino detectors are located underground at reactor flux weighted distances of 409~m (near position) and 1444~m (far position). As in Double Chooz and Daya Bay a 0.1~wt.\% Gd-LS was filled into cylindrical acrylic vessels with a volume of 18.6~m$^3$ in each detector. 

The composition of the RENO scintillator is very similar to the Daya Bay target liquid. Both use LAB as solvent, PPO (3~g/l) and bis-MSB as fluor combination and the carboxylic acid TMHA for the Gd loading. A difference between the Daya Bay and the RENO scinitllator production was that RENO used a two phase liquid-liquid extraction technique~\cite{RENOLS}. GdCl$_3$ was dissolved in ultrapure water and the pH value of the solution adjusted. Then it was mixed with a neutralized TMHA solution and LAB. The Gd is chemically transferred from the aqueous phase into the organic phase. Finally, the phases need to be separated again. In this way a stable Gd-LS with more than 10~m attenuation length (430~nm) and a primary light yield of about 8000~photons/MeV could be produced~\cite{RENOLS}, which corresponds to about 48\% of anthracene. 

\subsubsection{New generation experiments}
The Nucifer experiment is a very short baseline reactor neutrino experiment running at the Osiris research reactor in France~\cite{Nuc}. It was designed in the context of neutrino detectors for safeguard and non-proliferation. Moreover, it fulfills basic requirements to study the reactor neutrino anomaly~\cite{RNA} and sterile neutrinos. With a distance of about 7~m to the compact Osiris reactor core (70~MW thermal power) it can measure about 300 neutrinos/day in reactor ON phases with 850~liters of Gd-LS as Target. 

The Gd LS is based on the liquid used in Double Chooz. The same $\beta$-diketonate complex Gd(thd)$_3$ was used to dissolve the metal. Compared to the Double Chooz LS a higher Gd-concentration of 0.17~wt.\% was used. In this way background can be reduced by shortening the coincidence time window. Background in Nucifer is much higher than in Double Chooz, mainly due to the shallow depth and the reactor vicinity. The higher Gd-concentration, however, increases quenching and lowers the light yield. To compensate for this effect, the o-PXE to n-dodecane ratio in the LS was increased to about 50\% and more of the primary fluor PPO was added. In this way not only light yield, but also the pulse shape discrimination capability are improved.

There are several upcoming reactor neutrino experiments better suited and adapted to search for sterile neutrinos in the MeV range. Many of them as Stereo~\cite{LSNWP}, NEOS~\cite{NEOS} or Neutrino-4~\cite{Neu4} will use Gd-loaded LS for the antineutrino detection. Those will profit from the extensive R\&D work on Gd-loaded LS from previous experiments. In particular, the successful performance of the Gd-LS of Double Chooz, Daya Bay and RENO are setting a good basis for upcoming scintillator productions. In Stereo and NEOS, DIN is added to the scintillator at concentrations of 5-10\% to increase pulse shape discrimination capabilities. There are also plans to search for sterile neutrinos with $^6$Li doped LS as in PROSPECT~\cite{Pros}.

\subsection{Solar neutrinos}
In contrast to nuclear reactors which mainly produce electron antineutrinos, our Sun is a strong and pure source of electron neutrinos. The Sun creates its energy by the fusion of hydrogen to helium. The main fraction of the solar neutrinos are produced in the first step of the proton-proton chain reaction in which two protons produce deuterium, a positron and a neutrino (pp-neutrinos). The energy of these pp-neutrinos reaches only up to about 400~keV, which is a challenge for many detection techniques. In some branches of the pp chain higher energy neutrinos with more than 5~MeV energy are produced at lower rate. 

The first solar neutrino experiments could basically be classified in two detection techniques, one based radiochemical methods and the other using water Cerenkov detectors. The radiochemical experiments were the pioneering Homestake experiment from R.~Davis~\cite{Hom} and the gallium experiments GALLEX/GNO~\cite{Kae} and SAGE~\cite{SAGE}. In these experiments chlorine (Homestake) or gallium isotopes (GALLEX/GNO, SAGE) are converted into unstable argon or germanium nuclei. The few argon or germanium atoms produced in the CC neutrino reactions over several weeks of exposure are extracted from the multi-tons target material by chemical methods. The gaseous reaction products are filled into proportional counters and the rate of neutrino interactions is retrieved from the number of radioactive decays. 

The approach of kton water Cerenkov detectors was followed in Super-Kamiokande~\cite{SK} and SNO~\cite{SNO}. Here, elastic neutrino-electron scattering is used. In the SNO experiment several additional detection channels exist, which allow measurement of both the electron neutrino flux (CC reactions) and the total neutrino flux (NC reactions). In this way the longstanding ``Solar Neutrino Problem'' could be solved. With the SNO data it was shown that the observed deficit of detected solar electron neutrinos on Earth in Homestake, the Ga-experiments and SuperKamiokande is explained by conversion into the other neutrino flavors by the effect of oscillations~\cite{SNO}.     

Both technologies presented above have some drawbacks. They are either sensitive only at high energies (water Cerenkov), where the solar neutrino flux is low and the uncertainty on the predictions high, or they are lacking spectral or time information (radiochemical experiments). Both cases are solved by the use of a liquid scintillator detector. Several metal loaded LS detectors were proposed with various isotopes as neutrino target in the past. Nevertheless, the only liquid scintillator experiments measuring solar neutrinos so far are Borexino~\cite{Bor} and KamLand~\cite{Kamso}, which both use an unloaded pure LS. 

\subsubsection{Borexino}
The Borexino history started with first ideas on a liquid scintillator detector loaded with boron~\cite{Bor89}. The goal was to measure the CC and NC interactions of solar neutrinos on $^{11}$B nuclei to solve the solar neutrino puzzle. The CC reaction of interest was the inverse beta decay from $^{11}$B to $^{11}$C. In 26\% of the cases the $^{11}$C nucleus is left in an excited state. In this way the coincidence of an electron, a gamma from the $^{11}$C deexcitation and a final $\beta^+$-decay of $^{11}$C (20~min half-life) is expected. In the NC reaction the signal would be a single gamma from the an excited $^{11}$B state. The basic concept of comparing rates measured in CC and NC was similar to the later SNO-experiment. The scintillator technology offered the possibility of a ``live'' detector with low energy threshold and good energy resolution. It was reported that a B-loaded scintillator was tested, which contained 80\% trimethylborate (10\% boron content) and gave a light yield of 52\% of anthracene with an attenuation length $>10$~m~\cite{Bor89}.

Finally, due to the extraordinary radiopurity requirements on the LS for solar neutrino detection the Borexino Collaboration changed to a simpler binary LS system with PC as solvent and PPO (1.5~g/l) as single fluor. To achieve the ultra low background levels, the design of Borexino uses graded shielding, since then it is typical for common large LS neutrino detectors. At the center of concentric shells with increasing radiopurity there are 300~tons of LS~\cite{Bortech}. This inner volume is surrounded by another 1000~tons of liquid, a passive shield composed of PC and 5.0~g/l DMP (dimethylphthalate), a quencher for scintillation light.  More than 2000 eight-inch PMTs are mounted on a stainless steel sphere with 6.85~m radius. The detector design is completed by a water Cerenkov tank as outer veto detector. With this setup, Borexino succeeded in performing the first real-time spectral measurement of sub-MeV solar neutrinos starting with the monoenergetic $^7$Be-neutrinos of 0.86~MeV energy~\cite{Bor}. At a later stage the outstanding radiopurity of the LS even allowed the detection of the low energy pp-neutrinos~\cite{Borpp}.   

\subsubsection{Indium scintillators for solar neutrino detection}
\begin{figure}
\begin{center}
\includegraphics[scale=0.30]{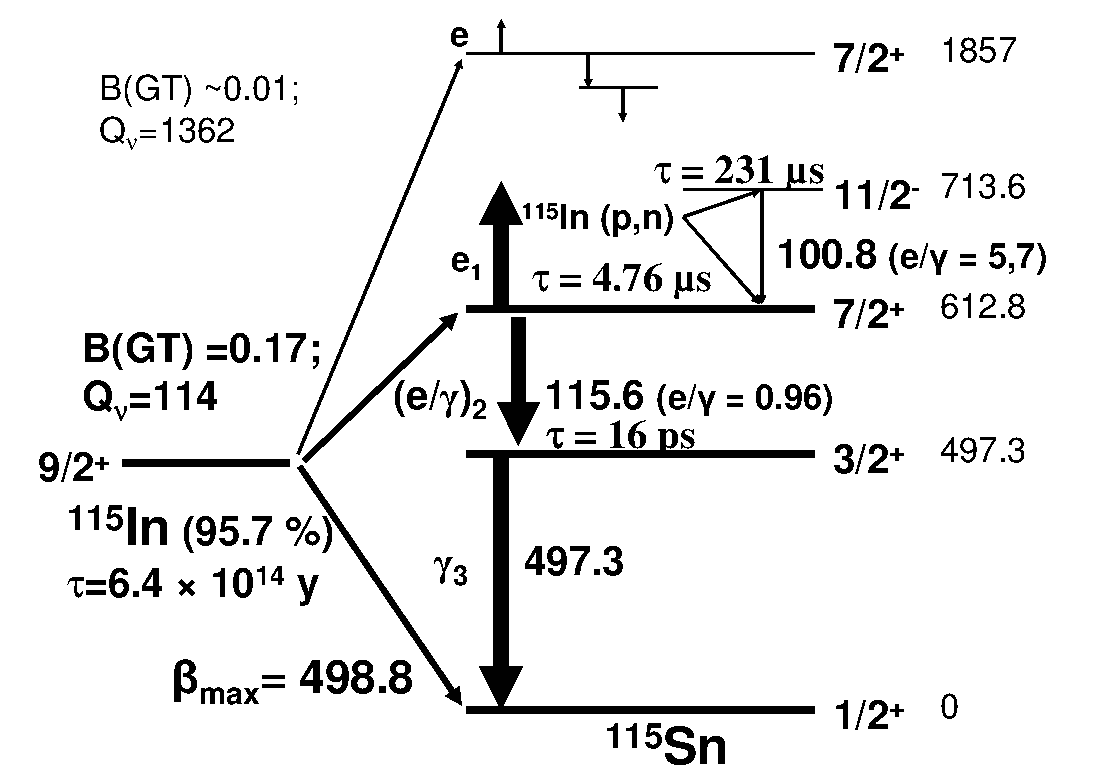}
\end{center}
\caption{The term scheme of the $^{115}$In-system is shown. Energies are given in keV. The energy threshold for the neutrino reaction is only 114~keV. In the reaction an excited state of $^{115}$Sn is populated which has a lifetime of 4.76$\mu$s and decays to the ground state under gamma emission.}
\label{Fig5}
\end{figure}

The goal of a real-time measurement of low energy solar neutrinos in the sub-MeV region triggered several studies on In-loaded LS over several decades~\cite{Ra76, PFE, PAY, SUZ}. Compared to detection techniques based on the elastic neutrino-electron scattering, In-loaded LS have the advantage of using the inverse beta decay of $^{115}$In to the 614~keV excited state of $^{115}$Sn as shown in Figure~\ref{Fig5}. The delayed coincidence signature of a prompt electron containing information about the neutrino energy, followed by a delayed (few $\mu$s) gamma cascade can be used for distinct background reduction. However, $^{115}$In is a $\beta^-$ emitter with a half-life of $4.4\cdot10^{14}$~a (0.25~Bq per gram of natural indium) and an endpoint energy of about 500~keV, which is a challenge for such detectors. For this reason and due to the additional difficulties of producing a stable and transparent liquid at high indium loading, all early attempts to develop an In-based LS detector were stopped. Nevertheless, the approach of using In-loaded LS for low energy solar neutrino detection was reconsidered in the beginning of this century and got integrated into the LENS activities, which originally proposed different isotopes.

\subsubsection{LENS}
The motivation for LENS was a measurement of the low energy pp-neutrino spectrum of the Sun. The majority of the neutrinos in the Sun are produced in this proton-proton fusion reaction. The pp-neutrino flux can be estimated from the luminosity of the Sun and is almost independent of solar model predictions. At the start of LENS three new candidates for low energy solar neutrino spectroscopy were proposed~\cite{Ra97}. These were the $\beta\beta$-decay candidates $^{176}$Yb, $^{160}$Gd and $^{82}$Se. They all have Q-values low enough to allow for the detection of pp-neutrinos via inverse $\beta$-decay. The early stages of the LENS R\&D phase mainly focused on $^{176}$Yb (Q-value of 301~keV). The excited state of the daughter nucleus $^{176}$Lu decays under emission of a 72~keV gamma with a half-life of 50~ns. The main challenges for this isotope were the rather short coincidence time and the rather low energy of the delayed gamma. For $^{82}$Se the situation in terms of timing and energy is even worse and Gd has other technical difficulties as the background from the alpha emitter $^{152}$Gd plus complicated level scheme of the daughter nucleus $^{160}$Tb. The main background sources identified for the Yb candidate isotope were $^{235}$U, $^{169}$Yb and $^{176}$Lu. Different Yb-loaded LS based on carboxylic acid systems were produced, but finally also Yb was dropped as target candidate and the collaboration decided to switch to $^{115}$In already proposed many years before LENS. 

Several groups inside the LENS collaboration were working on In-loaded LS with different carboxylic acids~\cite{Bara, Cha} or with $\beta$-diketonates~\cite{In1}. Two of the In-LS were tested along with unloaded reference cells in a low background environment at the Lens Low Background Facility (LLBF) in the Gran Sasso National Laboratory in Italy. The cells contained a carboylic acid version and a $\beta$-diketonate. In this setup it was even possible to measure the $^{14}$C content in the scintillator solvent in a cell of only 2~liters volume~\cite{LLBF}. For the indium carboxylates good results were obtained with isovaleric (5C) and 2-methyl valeric acid (6C)~\cite{Bara}. The candidate chosen for the LLBF test was 2-methyl valeric acid (MVA) due to its lower toxicity relative to the isovaleric acid. Indium has a high chemical reactivity to -OH groups even at low pH values. Part of the carboxylic -COO-groups are replaced by hydroxyl ions (-OH). Although hydrolysis can finally lead to instabilities and might cause the metal to come out of solution, some fraction of -OH groups helps in terms of light yield. The challenge is to keep solubility and avoid strong polymerization in the presence of -OH groups. In fact, in the prototype cell the indium was dissolved in the form of a polymeric hydroxycarboxylate of the form [In(OH)$_{2.2}$(2MVA)$_{0.8}$]$_n$~\cite{Dan08}. The -OH fraction can be tuned by pH adjustment in the liquid-liquid extraction. For an indium loading of about 54~g/l dissolved in PC at a fluor concentration of 4~g/l BPO (2-(4-biphenyl)-5-phenyloxazole) a light yield of about 54\% of anthracene could be achieved with an attenuation length of 3.9~m at 430~nm. Long term stability of the attenuation length could be confirmed over 6~years. The effective attenuation length of the liquid measured in the 1~m prototype cell was about 5~m~\cite{Bara}.

After the LLBF measurement the research on In-loqded LS for LENS continued in the US. Further improvements in terms of the optical properties of a similar indium carboxylate LS based on MVA were reported~\cite{Cha}. After a liquid-liquid extraction at pH 6.88 solid indium carboxylate is produced by vacuum distillation. The solid indium material can be directly dissolved in the scintillator solvent with an indium loading of 8~wt.\% and above. Even at such high loadings transparencies in the 10~m range at high light yields of 55\% were reported after production~\cite{Cha}. Similar to the LS described above, the indium complex contains -OH groups. The molecular formula found for the indium complexes is In(MVA)$_2$(OH). This composition is consistent with results from independent studies extracting the indium at a similar pH value around 6.9~\cite{Dan07}. After production there are mainly monomers, but after several months dimers, trimers and other oligomers could be identified in the LS. Those can cause a slight decrease of the LS transparency.  

\subsection{Neutrinoless $\beta\beta$-decay}
Attempts to measure the absolute magnitude of neutrino mass and test for the Majorana or Dirac nature of neutrinos via a rare radioactive decay process known as neutrinoless double beta decay ($0\nu\beta\beta$) have spanned decades from the early methods of radiochemical extraction to the recent technologies of bolometer, real-time scintillator (crystal, liquid, or sandwich-foil), semiconductor (e.g.~germanium) or hybrid LS+TPC~\cite{Ell}. These various detection schemes represent different approaches to distinguish differences in the summed electron energy spectrum near the decay endpoint, where backgrounds from internal and external radioactivity need to be greatly suppressed. Among these proposed experiments, KamLAND-Zen and SNO+ are the only scintillator-based detectors in which a $0\nu\beta\beta$ candidate is loaded in scintillant liquid to serve as detection medium and source.  

In principal, any double beta decay (DBD) isotope soluble in the liquid scintillator can serve as the source for $0\nu\beta\beta$ searches. Although the energy resolution of scintillator detectors will not be as good as competing solid-state or bolometric detectors, the large self-shielding volumes of high purity liquid and potential to load significant amounts of isotope could more than compensate for this deficiency. Furthermore, liquid scintillator detectors are readily scalable and have significant flexibility in the configuration in terms of loading levels and alternative deployments of $0\nu\beta\beta$ isotopes as well as in-line purifications.  	

\subsubsection{KamLAND-Zen}
The Kamioka Liquid Scintillator Antineutrino Detector (KamLAND) was built at the Kamioka Observatory, Japan, an underground neutrino detection facility to detect electron antineutrino oscillation. The KamLAND detector's outer layer consists of a 18~m stainless steel containment vessel equipped with about 2000 17-inch photo-multiplier tubes. Its second, inner layer, consists of a 13~m nylon balloon filled with a LS composed of thousand tonnes of 20~Vol.\% pseudocumene in 80~Vol.\% of n-dodecane loaded with PPO (1.52~g/L)~\cite{Kam08}. KamLAND-Zen is a successor of the KamLAND experiment that used the same detector to study double beta decay of $^{136}$Xe from a balloon deployed in the LS in summer 2011.

Xenon does not have long-lived radioactive isotopes other than $^{136}$Xe. The decay Q-value of the $^{136}$Xe$\rightarrow ^{136}$Ba transition is 2457~keV, among the highest values of $0\nu\beta\beta$ isotopes. Solubility of xenon gas in liquid scintillator is high, but decreases with increasing temperature. KamLAND-Zen uses 380~kg (3~wt.\%) of 91.7\% enriched $^{136}$Xe contained in a 1.54~m radius inner balloon, which was contaminated with Fukushima fallout, eventually leading to the leaching of nuclear-decay isotopes, particularly $^{110m}$Ag, in its scintillator.

Nevertheless, KamLAND-Zen set a limit for half-life of $1.9\cdot10^{25}$~y and measured the $2\nu\beta\beta$ lifetime as $(2.38\pm0.02(stat)\pm0.14(syst)\cdot10^{25})$~y, consistent with EXO’s results~\cite{KamZ}. KamLAND-Zen plans continued observations with 600-800~kg more enriched $^{136}$Xe and improved detector components including higher quantum efficiency photomultiplier tubes, light-collection mirror, and new, higher light-yield LS, LAB+PPO (2~g/l).

\subsubsection{Neutrinoless $\beta\beta$ decay with Neodymium}
For several reasons the rare earth element Neodymium (Nd) is a promising canidate for $0\nu\beta\beta$-decay searches. The isotope $^{150}$Nd (5.6\% natural abundance) has the shortest calculated half-life of all isotopes for a given effective Majorana neutrino mass~\cite{Elli}. Moreover, the Q-value for $^{150}$Nd is at 3.37~MeV, which is above the energy of typical backgrounds from natural radioactivity. Neodymium is rather inexpensive and can be deployed in large quantities into LS.

Such Nd-loaded LS were for example proposed and prepared in the context of SNO+~\cite{Chen}. Concentrations of Nd of more than 1\% have been tested. Most attempts of Nd loading were based on carboxylic acid versions. For a specific metal loaded LS with a concentration of 6.5~g/l Nd dissolved in PC a light yield of 75~\% compared to the unloaded scintillator version was achieved in a small sample and an effective optical attenuation length of 2~m was measured in a 2~liter cell~\cite{BarNd}.

The drawback of Nd are intrinsic absorption peaks in the PMT sensitive region of the spectrum. There is a transparent window at 370-415~nm, where attenuation length can be high ($> 5$~m). On the other hand at wavelengths from 415-500~nm several structures can be found in the absorption spectrum of Nd, which can not be removed by purification. Therefore, for a large SNO+ type detector light absorption at $\sim1$\% loadings is too high with common secondary wavelength shifters.

\subsubsection{SNO+}
SNO+ is a large liquid scintillator-based experiment located 2~km underground ($5890\pm94$~mw.e.) at SNOLAB, Sudbury, Canada. It reuses the Sudbury Neutrino Observatory detector, consisting of a 12~m diameter acrylic vessel to be filled with 780~tonnes of ultra-pure liquid scintillator. Designed as a multi-purpose neutrino experiment, the primary goal of SNO+ is a search for the neutrinoless double-beta decay. After initial R\&D work on $^{150}$Nd the focus is now on $^{130}$Te as a target. 

The isotope $^{130}$Te has the highest natural abundance among the double-beta decay isotopes and is of great interest as a $0\nu\beta\beta$ source owing to its low $2\nu\beta\beta$ rate, good optical transparency in the visible region, and high natural abundance (34.1\%, Q$=2529$~keV), which negates the need for costly enrichment. Yet it is not feasible to employ Te in LS using the conventional complexing-ligand, -COOH or -P=O, organometallic loading method due to its hydrophilic nature. Several attempts of Te-loaded LS have been made over the past few years, most Te-carboxylates or -diketones dissolved in liquid scintillators formed colored solutions, which are not suitable for a large production.

A new loading method using butanediol as complexing ligand for Te extraction at concentration levels of several percent into LAB is currently developed for SNO+. A superior optical transmission with reasonable light yield is achieved at 0.5~wt.\% of Te in LS, which is highly feasible for large scale detector. The diol-complexing method is now selected for SNO+ phase-I deployment. The details of Te-diol preparation will be presented in separate articles.

As mentioned above the water-based LS \cite{WLS15a} offers an alternate approach of loading metallic ions in aqueous phase directly into organic scintillator. Telluric acid, Te(OH)$_6$, is first dissolved in water and then loaded into LAB with the addition of a surfactant or a mixture of two surfactants. The Te-WbLS at 5 wt.\% has been prepared and stable for 1.5 years with better light-yield than that of Te-diol-LS. However to escape from color-quenching by self-absorption bands close to 400~nm and reduce the scattering effect induced by addition of surfactants, a secondary wavelength shifter added to the scintillator cocktail or an improved purification method of surfactants are required. 

In Phase I, the detector will be loaded with 0.5~wt.\% natural tellurium, corresponding to nearly 800~kg of $^{130}$Te, with an expected effective Majorana neutrino mass sensitivity in the region of 55-133~meV, just above the inverted mass hierarchy. Higher loading development is continuing, the possibility to deploy up to ten times more natural tellurium has been investigated, which would enable SNO+ to achieve sensitivity deep into the parameter space for the inverted neutrino mass hierarchy in the future. 

Additionally, SNO+ also aims to measure reactor antineutrino oscillations, low-energy solar neutrinos, geo-neutrinos, to be sensitive to supernova neutrinos, and to search for exotic physics. A first phase with the detector filled with water will begin soon, with the scintillator phase expected to start after a few months of water data taking. The Phase I is foreseen to start in 2017.

\subsubsection{Intrinsic and Cosmogenic Backgrounds}
The impact of background in a $0\nu\beta\beta$ experiment heavily relies on the energy resolution of the detectors. The solid-state or bolometric detectors have exceptional energy resolution that reduces the tail of the $2\nu\beta\beta$ spectrum from overlapping the region of interest of the $0\nu\beta\beta$ spectrum. Moreover, the contamination by for example the radioactive background from the 2.6~MeV $\gamma$ emanating from $^{206}$Tl is reduced. To compensate the deficiency of energy resolution of scintillator detectors ($3\sim4$\% at 2.6~MeV), high purity LS and $0\nu\beta\beta$ target are essential for the experiments to succeed.

Very low radioactivity of liquid scintillators (U at $10^{-18}$~g/g and Th at $10^{-17}$~g/g) has been reported by Borexino and KamLAND. KamLAND-Zen will continue to have highly pure $^{136}$Xe isotopes. Enrichment is achieved with centrifuged method, which moreover frees the material from other inherited gas contaminants. The balloon cleanness is improved as well (U/Th/K at $10^{-13}$~g/g), in addition to the removal of $^{110m}$Ag fallouts from its LS by distillation. For SNO+, tellurium compounds, as an element in the Group-VIA family, typically have relatively low actinide contaminations compared with lanthanide isotopes (e.g.~Nd). However, even though Te has low intrinsic U/Th backgrounds, the cosmogenic production of radioactive isotopes, such as sodium ($^{22}$Na), cobalt ($^{60}$Co), silver ($^{110m}$Ag), antimony ($^{124,126m}$Sb) and tin ($^{126}$Sn), could still contaminate the $0\nu\beta\beta$ region. Some of these cosmogenic isotopes, i.e.~Sn, with high hydrolysis constants can be removed along with U and Th by the self-scavenging method that was previously developed for gadolinium (Gd), neodymium (Nd), and applicable to other lanthanides via their feeding solutions~\cite{Yeh2010}. However, other cosmogenic isotopes such as Co or Ag are very water soluble, which implies that this scavenging method is not effective for their removals. Consequently an acid and thermal recrystallization method for Te compounds prior to its preparation of loading in LS is developed to cleanse these cosmogenic and radiogenic contaminants during the metal-loading process. This purification depletes U and Th and several cosmogenic isotopes from Te feedstock by a factor of $10^2 - 10^3$ in a single pass~\cite{HANS}. The process is also found to improve the optical transmission by removing traces of colored impurities. In addition to the scintillator-based experiments, this cleansing scheme has potential applications to the production of radiopure tellurium crystals for other rare-event experiments.

\newpage
\section{Conclusions}
The history of neutrino experiments is accompanied by intensive research and optimizations of metal loaded organic liquid scintillators. Various metals were considered in many different experiments, either to improve the detection efficiency of the neutrino interactions in the detector or as target material directly. Metal loaded liquid scintillators are of particular interest to search for neutrinos produced in nuclear reactors (e.g.~Gd, Li or Cd), the Sun (e.g.~In or Yb) or from $\beta\beta$-decay candidates (e.g.~Nd or Te).

For several decades the big challenge was to get the metal loaded into the organic liquid without affecting the performance of the scintillator. As neutrino interactions are rare processes the event rates in the experiments are typically low. This implies that detectors have to be large and need to run for several years to collect sufficient statistics. Therefore, the demands on the liquid properties are high transparency, radiopurity and long term stability.

In particular the stability was an obstacle in many experiments limiting the detector lifetime, causing delays due to extensive R\&D work or even preventing detector construction. Nowadays two rather mature methods to produce pure, transparent and stable metal-loaded liquid scintillators are available. One is based on the use of carboxylates for metal loading. This technology was already applied from the very beginning, but over time the procedures and material selections were more and more improved. The stable performance over several years of data collection in large scale neutrino detectors has been demonstrated recently. With this technology all basic requirements for neutrino detection can be met, including minor light quenching even at high loadings. A second approach, also applied successfully in currently running detectors, is based on $\beta$-diketonate chemistry with the main focus on chemical stability and radiopurity. In addition, the more recent loading technologies developed from quantum dots and water-based liquid scintillators expand the physics implications for scintillator detectors. With the gained knowledge the feasibility of future metal loaded liquid scintillator detectors in neutrino research seems promising and the base for further outstanding discoveries in the field is set.

\end{document}